\documentclass[aps,prd,12pt,superscriptaddress]{revtex4-2}
\usepackage{graphicx} 

\usepackage{comment}

\usepackage{caption}

\usepackage{subfig}

\usepackage{hyperref}

\usepackage{amsmath}

\usepackage{xcolor}

\usepackage[normalem]{ulem}

\begin{document}

\title{Radial perturbations of Ellis-Bronnikov wormholes in slow rotation up to second order}

\author{Bahareh Azad}
\email{bahareh.azad@uni-oldenburg.de}
\affiliation{Institute of Physics, University of Oldenburg, D-26111 Oldenburg, Germany}
\author{Jose Luis Bl\'azquez-Salcedo}
\email{jlblaz01@ucm.es}
\affiliation{Departamento de F\'isica Te\'orica and IPARCOS, Facultad de Ciencias F\'isicas, Universidad Complutense de Madrid, 28040 Madrid, Spain}
\author{Fech Scen Khoo}
\email{fech.scen.khoo@uni-oldenburg.de}
\affiliation{Institute of Physics, University of Oldenburg, D-26111 Oldenburg, Germany}
\author{Jutta Kunz}
\email{jutta.kunz@uni-oldenburg.de} 
\affiliation{Institute of Physics, University of Oldenburg, D-26111 Oldenburg, Germany}

\begin{abstract} 
We consider slowly rotating Ellis-Bronnikov wormholes and investigate their radial perturbations ($\mathrm{l}=0$), expanding up to second order in rotation.
We present the detailed derivations in the general case, including symmetric and non-symmetric wormholes.
The calculations show that the unstable mode present in the static case becomes less unstable with increasing rotation, until it reaches zero and then disappears.
This indicates that wormhole solutions may become linearly mode stable at sufficiently fast rotation.
\end{abstract}

\date{\today}

\maketitle

\newpage

\section{Introduction}

The prospects of obtaining traversable wormhole solutions have inspired much work in General Relativity and generalized theories of gravity.
In General Relativity traversable wormholes require either the presence of exotic matter \cite{Ellis:1973yv,Ellis:1979bh,Bronnikov:1973fh,Morris:1988tu,Morris:1988cz,Visser:1995cc} or of fermionic fields \cite{Blazquez-Salcedo:2020czn,Konoplya:2021hsm}, in order to violate the energy conditions.
On the other hand, in generalized theories of gravity this violation can arise directly from the gravitational fields as discussed, for instance, in \cite{Kanti:2011jz,Alcubierre:2017pqm}.

Among the numerous aspects of wormholes that have been considered, in particular, their observational signatures are of great importance.
In this respect already the earliest studies have considered gravitational lensing effects of wormholes
\cite{Cramer:1994qj,Safonova:2001vz,Perlick:2003vg,Nandi:2006ds,Abe:2010ap,Toki:2011zu,Nakajima:2012pu,Tsukamoto:2012xs,Kuhfittig:2013hva,Bambi:2013nla,Takahashi:2013jqa,Tsukamoto:2016zdu}.
Within the last decade and the associated observations of the EHT collaboration \cite{EventHorizonTelescope:2019dse,EventHorizonTelescope:2022wkp}
the investigation of wormhole shadows has attracted much interest
\cite{Bambi:2013nla,Nedkova:2013msa,Ohgami:2015nra,Shaikh:2018kfv,Gyulchev:2018fmd,Guerrero:2022qkh,Huang:2023yqd},
and likewise the study of wormhole accretion disks or of their corresponding quasi-periodic oscillations
\cite{Harko:2008vy,Harko:2009xf,Bambi:2013jda,Zhou:2016koy,Lamy:2018zvj,Deligianni:2021ecz,Deligianni:2021hwt}.

Depending on the respective wormhole solution and its associated environment, these observational signatures of wormholes may lead to quite unique predictions.
On the other hand the predictions for such wormhole properties may also mimic the corresponding predictions of the properties of black holes, making their observational identification more difficult 
\cite{Damour:2007ap,Bambi:2013nla,Azreg-Ainou:2014dwa,Dzhunushaliev:2016ylj,Cardoso:2016rao,Konoplya:2016hmd,Nandi:2016uzg,Bueno:2017hyj,Blazquez-Salcedo:2018ipc,Azad:2020ajs,Azad:2022qqn}.

One of the crucial desired properties of traversable wormholes is their stability \cite{Morris:1988cz}.
However, static spherically symmetric wormholes are haunted by a radial instability, as explicitly shown for the well-known Ellis-Bronnikov wormholes in General Relativity \cite{Shinkai:2002gv,Gonzalez:2008wd,Gonzalez:2008xk,Cremona:2018wkj,Blazquez-Salcedo:2018ipc} or for the wormholes in certain generalized gravity theories \cite{Cuyubamba:2018jdl}.
On the other hand, it has been conjectured, that rotation might stabilize wormholes \cite{Matos:2005uh}.
 
When Ellis-Bronnikov wormholes are set into rotation, the resulting spacetimes are only known perturbatively \cite{Kashargin:2007mm,Kashargin:2008pk} or numerically \cite{Kleihaus:2014dla,Chew:2016epf} despite efforts to obtain the rapidly rotating solutions in closed form \cite{Volkov:2021blw}.
We note though, that non-asymptotically flat rotating wormholes have been obtained in closed form \cite{Cisterna:2023uqf}, that correspond to Barcel\'o-Visser wormholes \cite{Barcelo:2000zf} in a swirling universe.

The fate of the radial instability has been investigated first for rotating wormholes in higher dimensions \cite{Dzhunushaliev:2013jja}.
Here the restriction to odd dimensions and equal magnitude angular momenta has provided a substantial simplification of the problem, leading to a system of ordinary differential equations. 
In this setting the unstable radial mode of static Ellis-Bronnikov wormholes becomes more stable with increasing rotation, while at the same time a second unstable radial mode arises from a zero mode in the static limit.
At a critical value of the rotation speed both unstable radial modes then merge and disappear.

Aiming at the goal of demonstrating the analogous behaviour for Ellis-Bronnikov wormholes in four dimensions, we have recently performed a perturbative analysis of the radial instability of these wormholes in second order in rotation \cite{Azad:2023iju}.
Our study has shown that, as in higher dimensions, the unstable mode becomes more stable with increasing rotation, while a second unstable mode arises from a zero mode in the static limit, as well.
Of course, being limited to slow rotation and thus allowing only for a quadratic dependence on the angular momentum, prevents the two modes from smoothly merging.
For the latter to happen we would have to resort to a more sophisticated numerical scheme.

Here we present the details of the calculations of the unstable radial-led modes in second order in rotation.
In particular, we present all formulae also for the non-symmetric wormholes, which have been omitted in \cite{Azad:2023iju}.
In section II we provide the theoretical setting, specifying the action and presenting the second order solutions for the slowly rotating wormholes.
We then consider the radial perturbations around these background solutions in section III, and also present the effective potential and the perturbation equations.
We explain the numerical method in section IV and discuss the results in section V. 
Section VI gives our conclusions.

\section{Theoretical setting}

\subsection{Action and field equations}

We consider General Relativity minimally coupled to a phantom scalar field $\Phi$.
The action reads
\begin{eqnarray}
			S &=& \frac{1}{16 \pi G}\int d^4x \sqrt{-g} 
		\Big[\mathrm{R} + 2 \partial_\mu \Phi \, \partial^\mu \Phi 
		 \Big] \, ,
   \label{eq:ellis} 
\end{eqnarray}
with curvature scalar $\mathrm{R}$ and Newton's constant $G$.
Variation of the action leads to the set of coupled field equations
\begin{eqnarray}
    \mathrm{R}_{\mu\nu} &=& -2 \partial_{\mu}\Phi\partial_{\nu}\Phi \, , 
    \label{m_eqs}  \\
    \partial_{\mu}\partial^{\mu}\Phi &=& 0 \, .
    \label{s_eq}
\end{eqnarray}

\subsection{Static background solutions}

The static background solution for Ellis-Bronnikov wormholes is given by
\begin{eqnarray}
ds^2 &=& -e^{f} dt^2 
+ e^{-f} dr^2 + {e^{-f}} R^2 \left(
d\theta^2+\sin^2{(\theta)}d\varphi^2 \right) \ ,
\label{metric_b}
\end{eqnarray}
with metric function $f(r)$
\begin{eqnarray}
    f(r) = \frac{{C}}{r_0} \left(\tan^{-1} \! \left(\frac{r}{r_0}\right)-\frac{\pi}{2}\right) \ ,
\end{eqnarray}
asymmetry parameter $C$, and $R^2=r^2+r_0^2$.
The static background scalar field $\Phi$ is given by
\begin{eqnarray}
  \phi(r)= \frac{Q_0 f}{C} \ ,
\end{eqnarray}
with 
$Q_{0} = \sqrt{C^2/4 + r_0^2}$.

Ellis-Bronnikov wormholes are asymptotically flat.
This is, however, only obvious for $r \rightarrow +\infty$, since the metric function $f$ tends to zero there, $f \rightarrow 0$. 
For $r \rightarrow -\infty$ asymptotic flatness becomes only apparent after a coordinate transformation
 \begin{equation} \label{coor_trans}
  \bar{t} = e^{-\frac{C \pi}{2 r_0}} t \,, \quad \bar{r} = e^{\frac{C \pi}{2 r_0}} r \, ,
 \end{equation}
unless $C=0$.
In that case the wormhole is symmetric with respect to reflection of $r \to -r$ at the throat $r=0$.
 
The parameter $C$ not only determines the asymmetry of the wormhole, but it is also associated with its global charges, that can be read off from the asymptotic expansions of the solution. 
For $r \to +\infty$ one then finds the wormhole mass $M_0 = C/2$ and the scalar charge $Q_{0}$.
In the symmetric case $C=0$, the static wormhole is massless, whereas in the non-symmetric case $C \neq 0$, the wormhole carries mass.
The parameter $r_0$ is associated with the size of the throat.
For $C=0$, the throat is located at $r=0$, while for $C\neq 0$ it is shifted to one of the sides as seen by determining the minimal surface.
Finally we recall, that there is a relation between wormhole solutions with positive values of the asymmetry parameter $C$ and those with a negative $C$,
\begin{eqnarray}
f(r,C) = f(-r,-C) - \frac{\pi C}{r_0}, \\
\phi(r,C) = -\phi(-r,-C) - \frac{\pi Q_0}{r_0} \,.
\end{eqnarray}

\subsection{Background solutions up to second order in rotation}

We now present the background solution up to second order in rotation for a finite asymmetry parameter, $C > 0$. Denoting the expansion parameter $\epsilon_r$,  the metric reads
\begin{eqnarray}
ds^2 &=& -e^{f}\left[1+\epsilon_r^2 2\left(h_0(r)+h_2(r)P_2(\theta)\right)\right]dt^2 
+ e^{-f}\left[1+\epsilon_r^2 2\left(b_0(r)+b_2(r)P_2(\theta)\right)\right]dr^2 \nonumber \\ 
&+& 
{e^{-f}} R^2
\left[1+\epsilon_r^2 2\left(k_0(r)+k_2(r)P_2(\theta)\right)\right]
\times 
\left[
d\theta^2+\sin^2{(\theta)}\left[d\varphi-\epsilon_rw(r)dt\right]^2 
\right]
\ ,
\label{metric_1}
\end{eqnarray}
with Legendre polynomial $P_2(\theta) = \left(3\cos^2{(\theta)}-1\right)/2$.
The phantom scalar field up to second order in rotation is given by
\begin{eqnarray}
\Phi = \phi(r) + \epsilon_r^2 \left(\phi_{20}(r)+\phi_{22}(r)P_2(\theta)\right) \ .
\label{scalar_1}
\end{eqnarray}

In our calculations, we apply the gauge $k_0=0$, and redefine $k_2=h_2-\nu_2$.
Thus we are left with two sets of background functions, $\mathcal{P}_0=\{h_0,{b_0},\phi_{20}\}$, and $\mathcal{P}_2=\{{h_2}, b_2, \nu_2, \phi_{22}\}$.
These two sets decouple.
Radial perturbations only require the $\mathcal{P}_0$ functions.
Nonetheless, for completeness we provide in the following the solutions to the $\mathcal{P}_2$ functions as well.
These would be necessary for a calculation of the quasinormal modes in second order in rotation for non-radial perturbations \cite{Blazquez-Salcedo:2022eik}.

\paragraph{First order ($C > 0$):}
To first order a single background function is present, $w(r)$, which reads
\begin{eqnarray}
    w(r)= \frac{3J}{2C(C^2+r_0^2)}
    \left[
    1 - \left(
    1+2C\frac{C+r}{R^2}
    \right)e^{2f}
    \right] \, . \, \, \, \, \, \, \, \,
    \label{wr}
\end{eqnarray}
Here $J$ is the angular momentum of the wormhole solution, read off for $r \to \infty$.

\begin{figure}
		\centering
    		\includegraphics[width=0.3\textwidth,angle=-90]{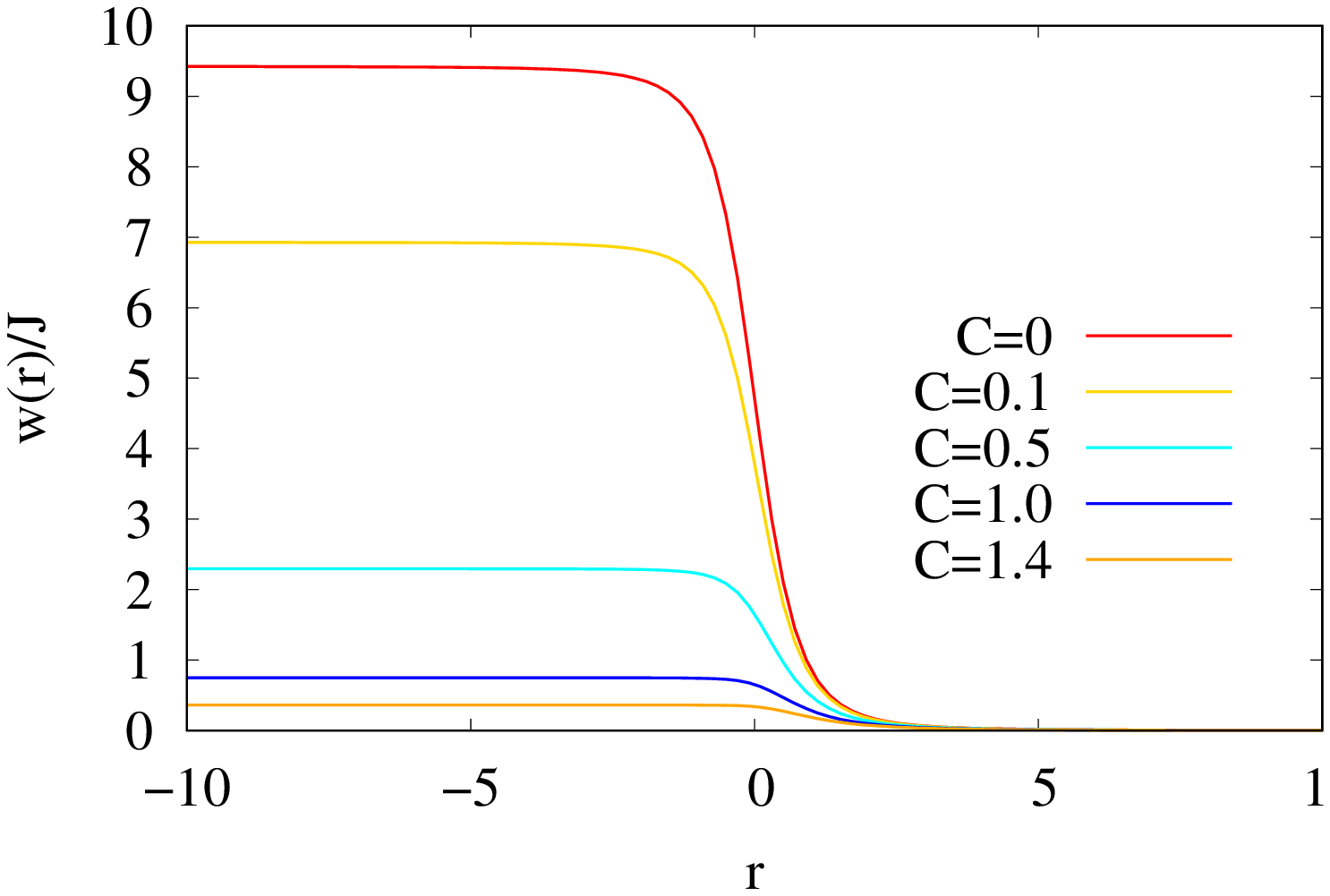}
\includegraphics[width=0.3\textwidth,angle=-90]{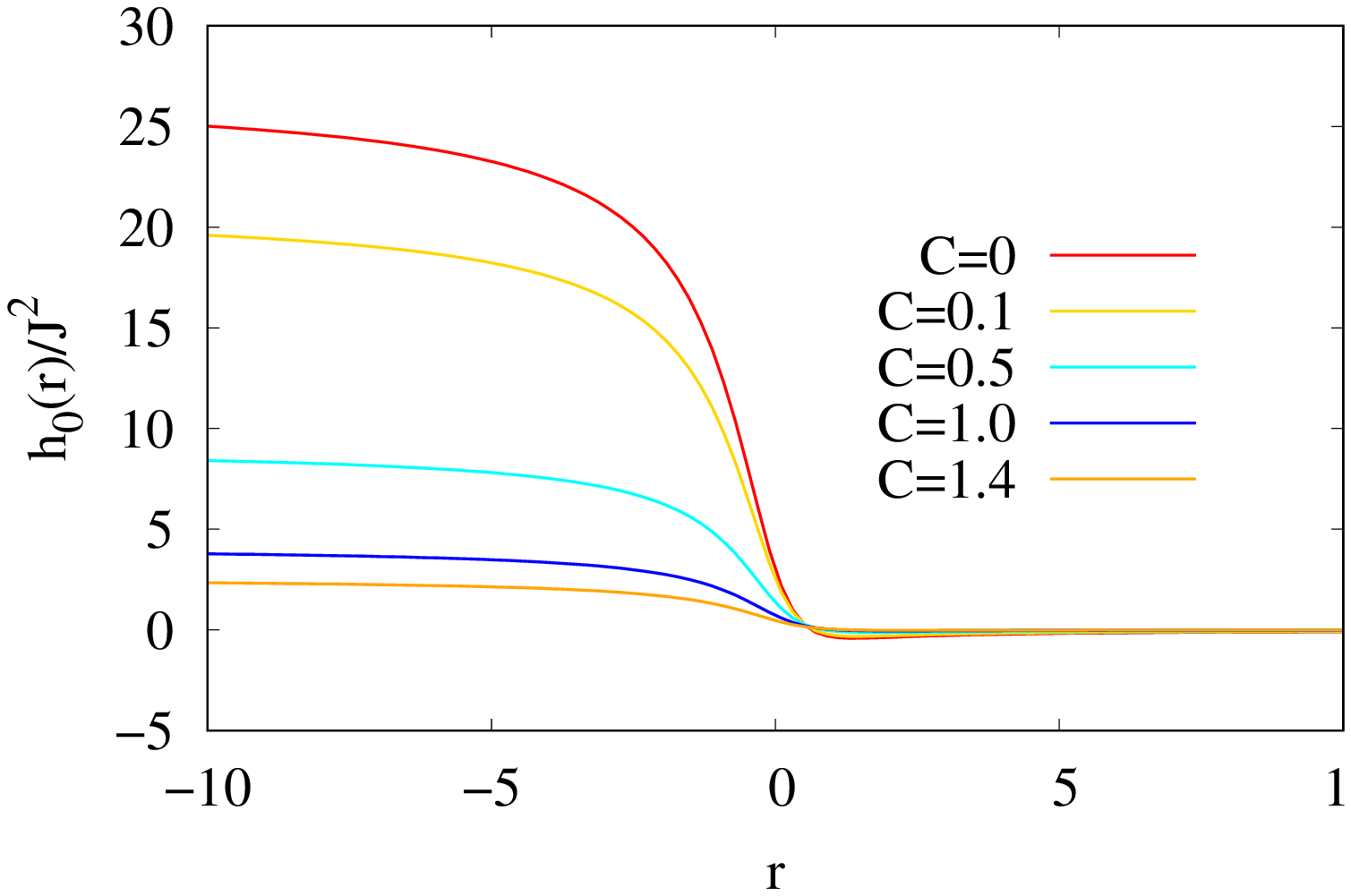}
\includegraphics[width=0.3\textwidth,angle=-90]{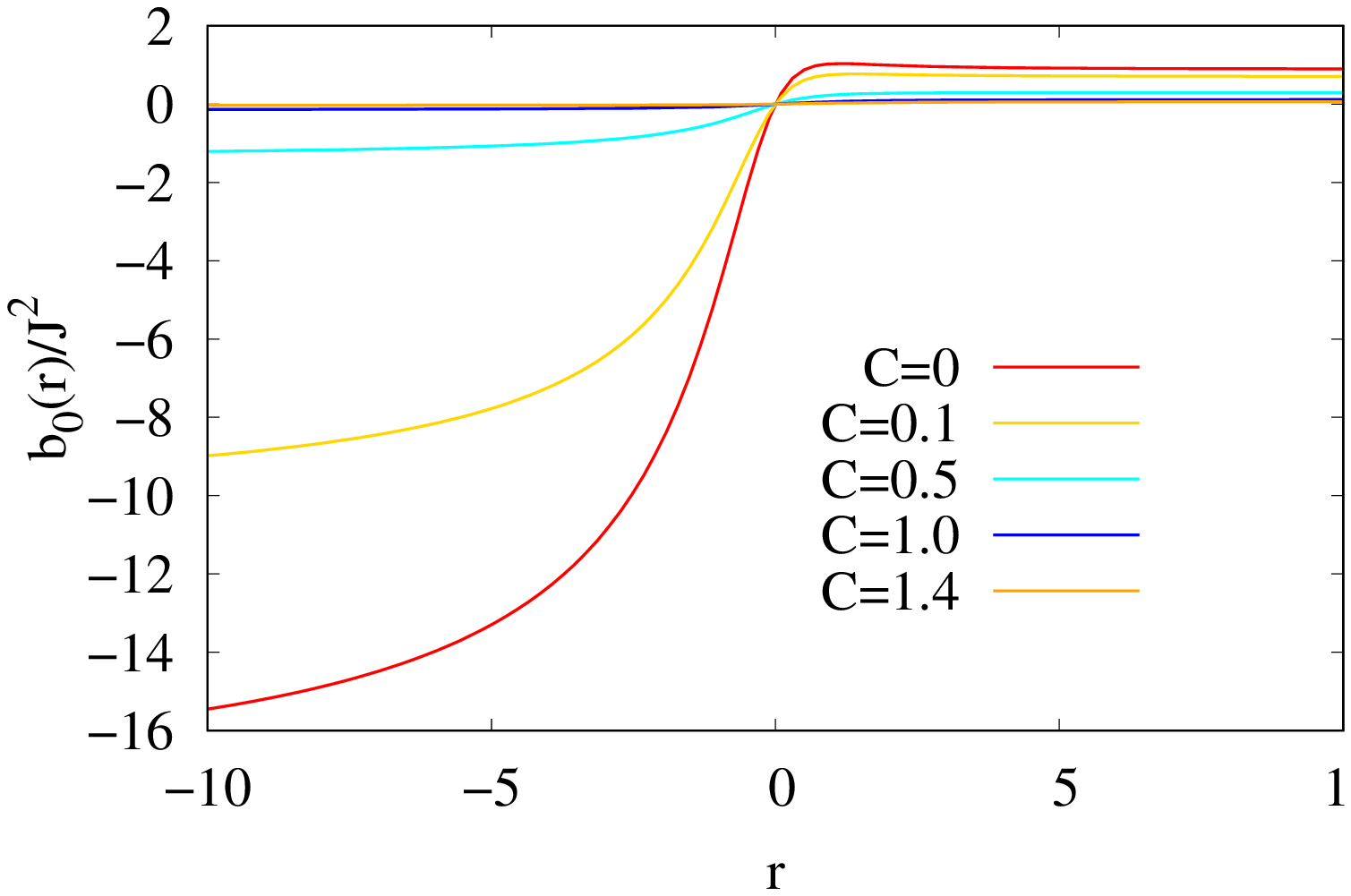}
\includegraphics[width=0.3\textwidth,angle=-90]{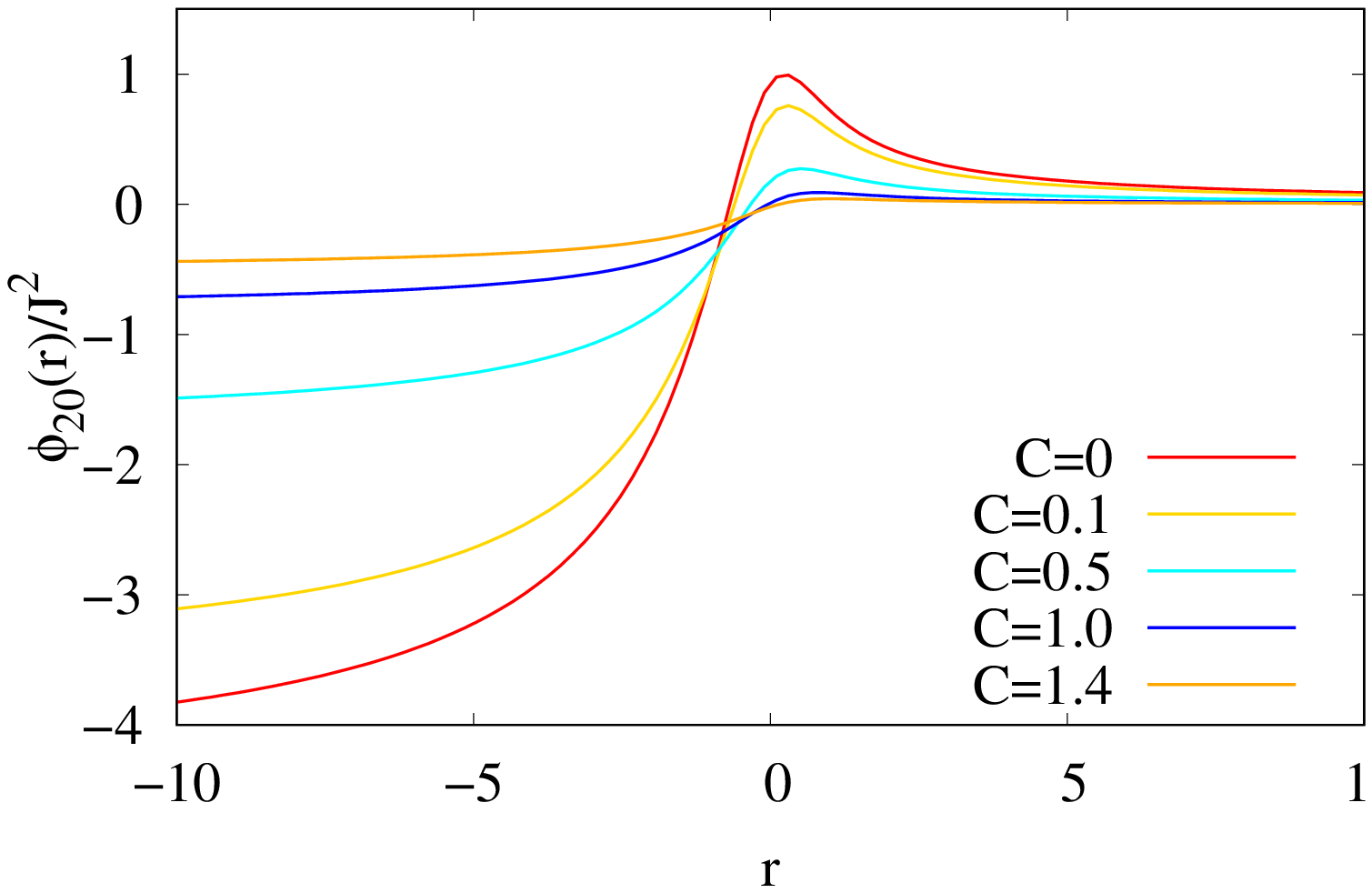}
  \caption{Scaled first order background function $w(r)/J$, eq.~(\ref{wr}),
  second order background functions $h_{0}(r)/J^2$, eq.~(\ref{h0_withc}), $b_{0}(r)/J^2$, eq.~(\ref{b0_withc}), and $\phi_{20}(r)/J^2$, eq.~(\ref{phi20_withc}), vs radial coordinate $r$ for several values of the asymmetry parameter $C$ for $r_0=1$. 
  Scaling parameter is the angular momentum $J$. } 
		\label{Fig_bckgrdP0_r}
	\end{figure}

\paragraph{Second order ($\mathcal{P}_0$, $C > 0$):}
We obtain for the functions in the set $\mathcal{P}_0=\{h_0,{b_0},\phi_{20}\}$ the solutions
\begin{eqnarray}
  h_0(r) &=& \frac{ 3 J^2 }{ \alpha}
  \big\{
  -17 R^2   e^{-\frac{C}{r_0}\left(\pi-2 \tan^{-1} \left(\frac{C}{2 r_0}\right)\right)}
  \big(C^4-\frac{2}{17} r_0^2 (C^2 -4r_0^2 )\big) 
  \nonumber 
  \\ && \times
 \big(-8r Q_0^2 \tan^{-1} \left(\frac{r}{r_0}\right) + (\pi r - 2 r_0)C^2 + 4\pi r r_0^2 \big) 
  \nonumber 
  \\ 
 & +& 4 Q_0^2  
  \big[ e^{2f}  
\alpha_1
  \big(C^2(-C^3+2rC^2+(6r^2-r_0^2)C+ 5r^3+rr_0^2) + 2rr_0^2 R^2\big)
     \nonumber 
  \\ 
 & +& 20 R^2 \left(C^2+\frac{2r_0^2}{5}\right) 
  \big( 2 r Q_0^2 \big(\tan^{-1} \left(\frac{C}{2r_0}\right) - \tan^{-1} \left(\frac{r}{r_0}\right) \big) 
  + Cr_0 \left(r-\frac{C}{2}\right) \big)
  \big]
  \big\} \, ,
  \nonumber 
  \\
  \label{h0_withc}
\end{eqnarray}
\begin{eqnarray}
    b_0(r) &=&
 -
 \frac{51 J^2 r r_0 R^{2}}{ \left(r -\frac{{C}}{2}\right) {C} \alpha}
 \, {e}^{\frac{ {C}   }{{r_0}} (2 \tan^{-1} \left(\frac{{C}}{2 r_0}\right)- \pi) + f}
\nonumber 
 \\ 
& \times &
 \big\{ 
\big(4 Q_0^{2} ({C} r +2 r_0^{2}) f -8 r {C} r_0^{2}+{C}^{4}+8 r_0^{2} {C}^{2} \big) 
\left({C}^{4}-\frac{2}{17} r_0^{2} {C}^{2}+\frac{8}{17} r_0^{4} \right) 
\nonumber 
\\ 
&-& \frac{10 Q_0^{2}}{17}
[
-8 Q_0^{2}
{C} \tan^{-1}  \left(\frac{{C}}{2 r_0}\right)
\big(\frac{{ e}^{3 f }}{5} 
\big(\beta 
- 
2 (r_0 RC)^{2} 
+6 r r_0^{2} R^{2} {C}
-4 r_0^{4} r^{2}-4 r_0^{6} \big) 
\nonumber 
\\    
&+&\left({C}^{2}+\frac{2 r_0^{2}}{5} \right) {\mathrm e}^{f } R^{2} ({C} r +2 r_0^{2}) \big) 
+\frac{1}{5}
\big(
\beta +
8  r_0^{2} R^{2} C  {e}^{3 f } (
-   {C}^{2}+ 3   r{C} 
-2  r_0^{2}  )
 ( \pi  Q_0^{2}- {C} r_0 )
\big)
\nonumber 
\\ 
&+& 
4 Q_0^{2} {e}^{f } R^{2}
\big(2{r_0}f( {C} r  +2 r_0^{2})  + C ((\pi  r +2  {r_0} ) {C} +2 r_0( \pi  r_0- 2r ) )  \big) 
\big({C}^{2}+\frac{2 }{5} r_0^{2} \big) 
]
\big \} \, ,
\label{b0_withc}
\end{eqnarray}
where
\begin{eqnarray}
    \beta = {C}^{3} ( -3 {C}^{3}+(2 r^{2}-7 r_0^{2}) {C}+ 3r(r^{2}- r_0^{2}) ) \, ,
\end{eqnarray}
and
\begin{eqnarray}
    \phi_{20}(r)
    &=& - \frac{ 6 J^2}{f \alpha}
    \nonumber
    \\ &\times&
\big    \{
    -34 (({C} r +2 r_0^{2}) f +{C}^{2}) \left({C}^{4}-\frac{2}{17} r_0^{2} {C}^{2}+\frac{8}{17} r_0^{4} \right) r_0 R^{2} 
    { e}^{-\frac{{C} } {r_0}(-2 \tan^{-1} \left(\frac{{C}}{2 r_0}\right)+\pi )} 
     \nonumber 
     \\ 
    &+& 4 Q_0^{2}
    \big[-8 Q_0^{2} {C} 
    -R^{2} (-3 {C}^{2}+{C} r -r_0^{2})
    \tan^{-1}  \left(\frac{{C}}{2 r_0}\right)
     \nonumber 
\\ 
    &+& { e}^{2 f } \big(-{C}^{2} (C+r)^2 +  r R^2 {C} -R^{2} r_0^{2} \big) 
     \big(
     \tan^{-1} \left(\frac{{C}}{2 r_0}\right)
     +
    4 C (  \pi  Q_0^{2}- {C} r_0 )
     \big)   
    \nonumber \\ 
    &-& R^{2} \big(-10r_0 f \left({C}^{2}+\frac{2 }{5} r_0^{2} \right)  ({C} r +2 r_0^{2})  + C(-3 {C}^{4} \pi +(\pi  r +2 r_0 ) {C}^{3}
      \nonumber
    \\ 
    &-& (13 \pi  r_0 
    +4 r ) r_0 {C}^{2}
     +4 \pi  r_0^{2} (rC - r_0^2)
   )  \big)  \big]
    \big\} \, ,
    \label{phi20_withc}
\end{eqnarray}
where
\begin{eqnarray}
 \alpha =    64 
 \left(r -\frac{{C}}{2}\right) \big( Q_0 {C} R ({C}^{2}+r_0^{2}) \big)^{2} \alpha_1
  \, ,
\end{eqnarray}
and
\begin{eqnarray}
    \alpha_1 = 
    -2 Q_0^2 \tan^{-1} \left(\frac{{C}}{2 r_0}\right)+
\pi Q_0^2 - 4{C} {r_0} \, .
\end{eqnarray} 
The background functions are illustrated in Figure \ref{Fig_bckgrdP0_r} for several values of the asymmetry parameter $C$.
We note that in the symmetric case $C=0$, the $\mathcal{P}_0$ functions are simpler.
They can be found in \cite{Azad:2023iju}.

\begin{figure}
		\centering
    		\includegraphics[width=0.3\textwidth,angle=-90]{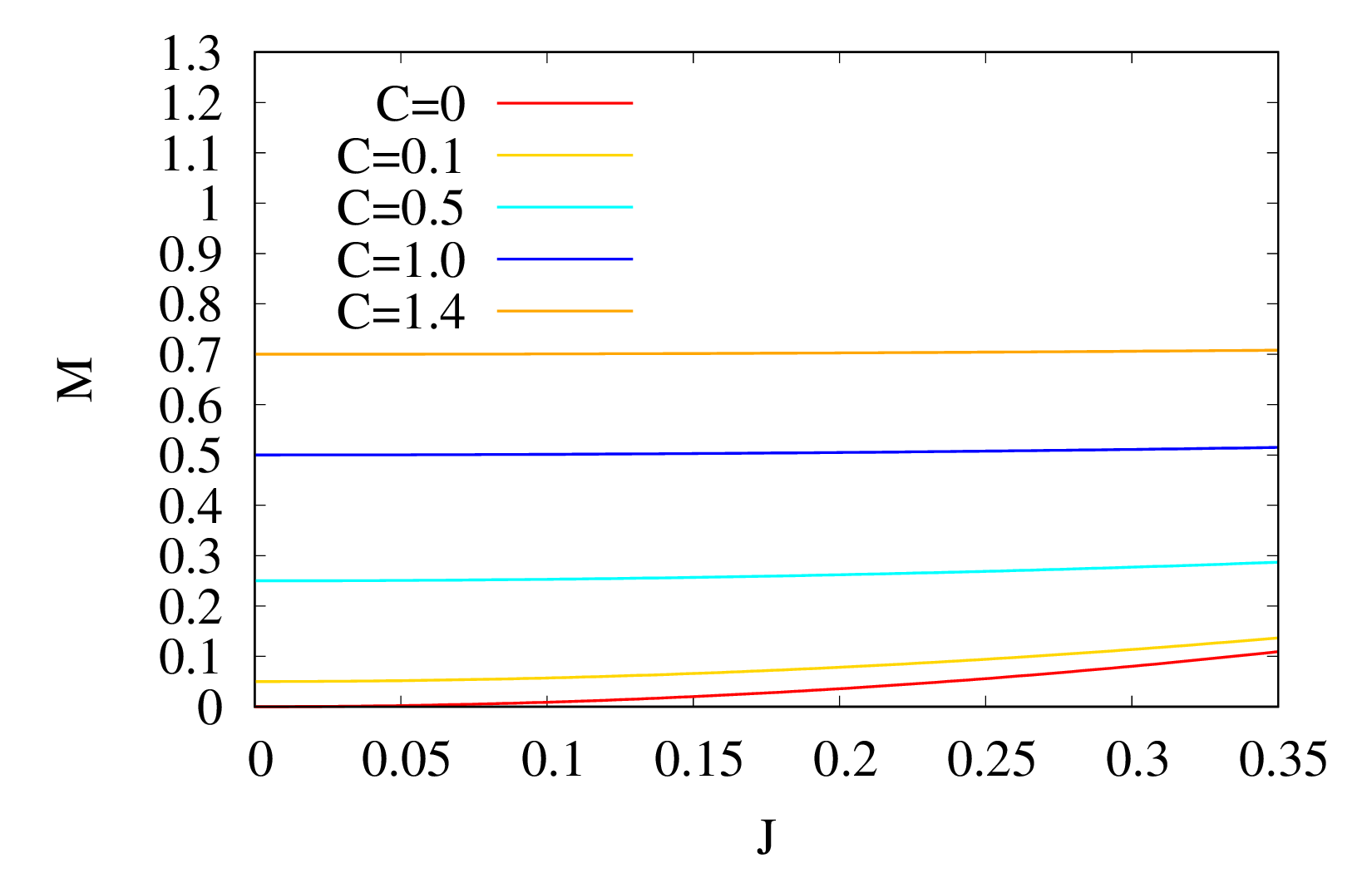}
\includegraphics[width=0.3\textwidth,angle=-90]{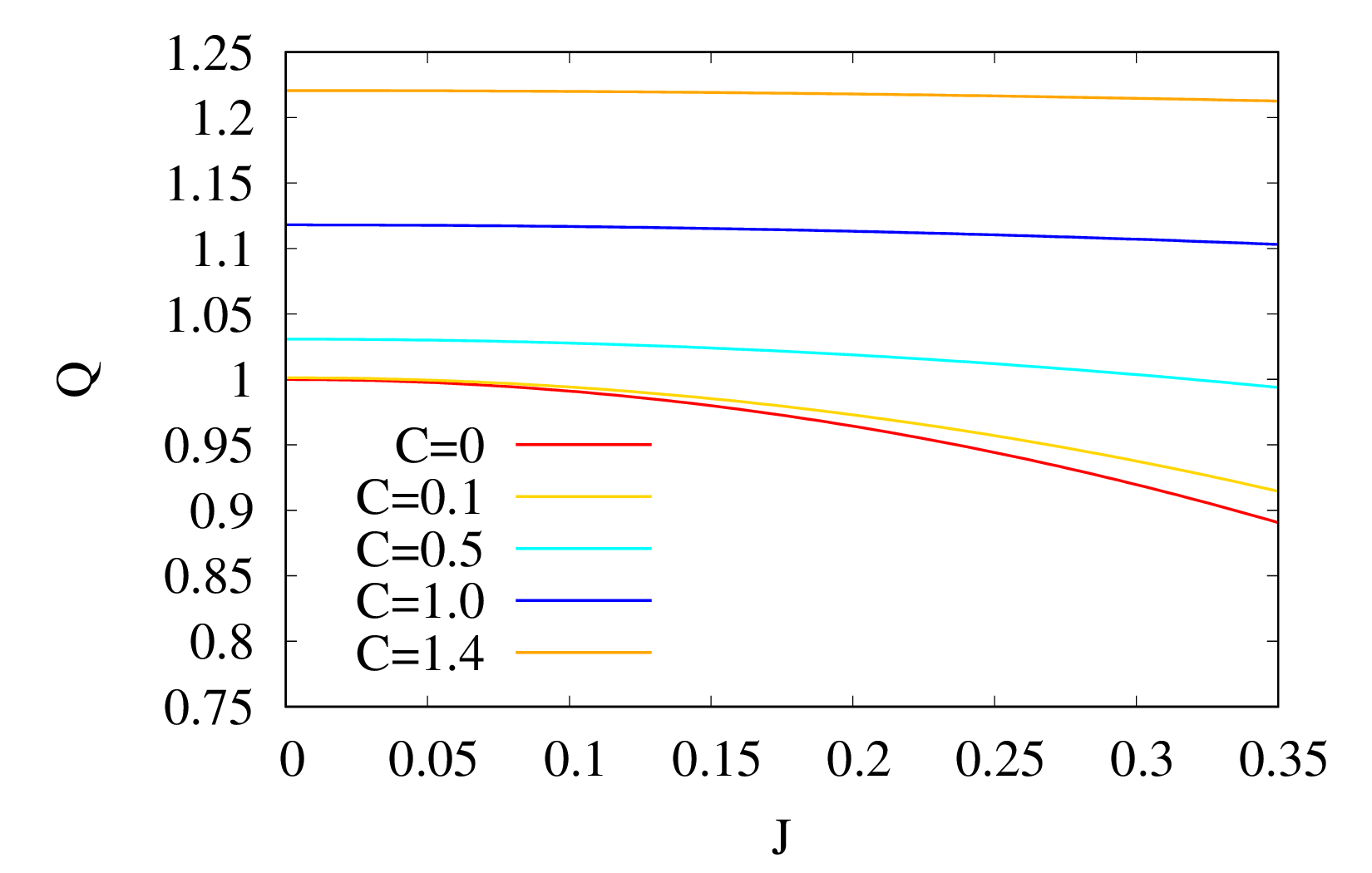}
  \caption{Mass $M$, eq.~(\ref{Mass}) (left), and scalar charge $Q$, eq.~(\ref{charge}) (right) at positive infinity vs angular momentum $J$ for several values of the asymmetry parameter $C$ for $r_0=1$. } 
\label{Fig_MQ}
	\end{figure}

To second order in rotation, we obtain a correction to the mass,
\begin{eqnarray}
{M} = M_0 + \epsilon_r^2\Delta M \, .
\label{Mass}
\end{eqnarray}
Evaluation of the correction for $r \to \infty$ yields
\begin{eqnarray}
 \Delta M &=&   \frac{12 J^2 R^2 (r-C/2)}{\alpha} \big\{
2 r_0^3 e^{-\frac{C}{r_0}(\pi - 2\tan^{-1} (C/2r_0))}
 (17C^4 - 2C^2 r_0^2 + 8 r_0^4) 
\nonumber 
  \\ &&   
+ 4Q_0^2
\big(
-2C (C^4+5r_0^2C^2+4r_0^4) \tan^{-1} \left(\frac{C}{2r_0}\right)
\nonumber 
  \\ &&   
+\pi C^5 -4r_0 C^4 + 5  \pi r_0^2C^3 
-14 r_0^3 C^2 + 4\pi r_0^4 C - 4 r_0^5
\big)
    \big\} \, .
    \label{Delta_M}
\end{eqnarray}
Analogously, the scalar charge also receives a correction,
\begin{eqnarray}
Q = Q_0 + \epsilon_r^2\Delta Q \, ,
\label{charge}
\end{eqnarray}
where $\Delta Q=-\Delta M$.
The mass $M$ and the scalar charge $Q$ are illustrated in Figure \ref{Fig_MQ} for several values of the asymmetry parameter $C$.

\paragraph{Second order ($\mathcal{P}_2$, $C = 0$):} 
Setting $C=0$, we find for the set of functions in $\mathcal{P}_2=\{{h_2}, b_2, \nu_2, \phi_{22}\}$
\begin{eqnarray}
    h_2^{(C=0)}(r) &=& \frac{3J^2}{r_0^5} \big\{
-\frac{1}{r_0} (3r^2+r_0^2) \left(\tan^{-1} \left(\frac{r}{r_0}\right)\right)^2
-6r \tan^{-1} \left(\frac{r}{r_0}\right)
\nonumber 
\\ 
&+& \frac{1}{4r_0R^2} \big(\pi^2r^2(3r^2+4r_0^2)+r_0^2(\pi^2r_0^2-12r^2-8r_0^2)\big)
    \big\}  \, ,
\end{eqnarray}
\begin{eqnarray}
    b_2^{(C=0)}(r) &=&
    \frac{3J^2r}{r_0^5} \big\{
\frac{1}{r_0} (3r^2+r_0^2) \left(\tan^{-1} \left(\frac{r}{r_0}\right)\right)^2
+6r \tan^{-1} \left(\frac{r}{r_0}\right)
\nonumber 
\\ 
&-& \frac{1}{4r_0R^4} \big(\pi^2r^2(3r^4+7r^2r_0^2+5r_0^4)
\nonumber 
\\ 
&+& r_0^2(\pi^2r_0^4-12r^4-20r^2r_0^2-16r_0^4)\big)
    \big\}  \, ,
\end{eqnarray}
\begin{eqnarray}
    \nu_2^{(C=0)}(r) =
    -\frac{3J^2}{R^4}  \, ,
\end{eqnarray}
\begin{eqnarray}
\phi_{22}^{(C=0)} (r) &=&
-\frac{6J^2r}{r_0R^4 (\pi-2\tan^{-1}(r/r_0))} \, .
\end{eqnarray}

\paragraph{Second order ($\mathcal{P}_2$, $C > 0$):} 
For $C>0$, we find
\begin{eqnarray}
h_2(r) &=& \frac{9J^2}{4C^2
R^4\pi(C^2 + r_0^2)^2} \nonumber \\
& \times& \Big[ -2\pi e^{2f} 
\Big( r^6 + 2Cr^5 + ( \frac{11C^2}{6} + \frac{7r_0^2}{3} )r^4 + \left( C^3 + 4Cr_0^2 \right)r^3 
\nonumber \\
&+& ( \frac{1}{3}C^4 + 3C^2r_0^2 + \frac{5}{3}r_0^4 )r^2  
- C \left( C^4 - 2C^2r_0^2 - 6r_0^4 \right)\frac{r}{3} - \frac{C^6}{3} + \frac{7C^2r_0^4}{6} + \frac{r_0^6}{3} \Big) 
\nonumber \\
& +& R^2 \Big(
\chi_1 (e^{-2C\pi/r_0}+1)
    + \chi_2(-e^{-2C\pi/r_0}+1)
\Big)
 \Big] \, ,
\end{eqnarray}
where
\begin{eqnarray}
    \chi_1 &=&  \pi r^4 
    - \frac{\pi}{6}(C^2 - 8r_0^2)r^2
    - \frac{\pi}{6}r_0^2(C^2 - 2r_0^2) \, ,
    \\
    \chi_2 &=&
    2 R^2 (r^2 - \frac{C^2}{6} + \frac{r_0^2}{3})\tan^{-1}(\frac{r}{r_0})
    + 2r^3r_0 + (2r_0^3 - \frac{1}{3}C^2r_0)r \, ,
\end{eqnarray}
\begin{eqnarray}
b_2(r) &=& \frac{9re^{-2C\pi/r_0}J^2}{2C^2 R^4 \pi(C^2 + r_0^2)^2} \nonumber \\
&\times& \Big[
-\frac{1}{2} R^2
 \Big( 
 \chi_1
    ( e^{\left(C(2\tan^{-1}(r/r_0) + 3\pi)\right)/2r_0}
    + e^f)
+
\chi_2
     ( e^{\left(C(2\tan^{-1}(r/r_0) + 3\pi)\right)/2r_0}
    - e^f)
    \Big)
\nonumber\\
&+& 
e^{\frac{C(6\tan^{-1}(r/r_0) + \pi)}{2r_0}}\pi
\big( r^6 + 2Cr^5 + (\frac{11C^2}{6} + \frac{7r_0^2}{3})r^4 + (C^3 + 4Cr_0^2)r^3 
\nonumber\\
&+& (\frac{1}{3}C^4 + 3C^2r_0^2 + \frac{5}{3}r_0^4)r^2 
- \frac{Cr}{3}
(C^4 - 2C^2r_0^2 - 6r_0^4) 
+ \frac{r_0^6}{3} + C^6 + \frac{5C^2r_0^4}{2} + \frac{8C^4r_0^2}{3} \big)
 \Big] \, , 
 \nonumber  \\
\end{eqnarray}
\begin{eqnarray}
\nu_2(r) &=&
-\frac{9J^2}{4\pi
R^4 C(C^2 + r_0^2)^2}
\Big(-2 \pi e^{2f} 
\Big[ r^5 + 2Cr^4 + 
2\left(C^2 + r_0^2\right)
(r^3
+ \frac{1}{3} Cr_0^2)
\nonumber\\
&+& \frac{2}{3}r^2 \left(2C^3 + 5 Cr_0^2\right) 
+
\left(\frac{2}{3}C^4 + \frac{8}{3}C^2r_0^2 + r_0^4\right)r 
\Big]
+
R^2 
\Big[
 \pi r R^2 
(e^{-2C\pi/r_0} +1)
\nonumber \\
&+& 
(2rR^2
\tan^{-1}\left(\frac{r}{r_0}\right) 
+ 2r^2r_0 + \frac{4r_0^3}{3} )
(-e^{-2C\pi/r_0} +1)
\Big]
\Big)\, ,
\end{eqnarray}
\begin{eqnarray}
\phi_{22}(r) &=& \frac{3r_0J^2}{2 \pi
R^4 (C^2 + r_0^2)^2
(-2r_0f)} \nonumber\\
&\times&
\Big[ -2\pi e^{2f}
( r_0^4 + (4C^2 + 4Cr + 2r^2)r_0^2 + r^4 + 2r^3C + 2C^2r^2 + 2C^3r + 2C^4 )  \nonumber\\
&+& R^2 
\Big( 
 \pi R^2   (e^{-2C\pi/r_0} +1) 
+ (2R^2 \tan^{-1}(\frac{r}{r_0})  + 2rr_0)
(-e^{-2C\pi/r_0} +1)
\Big) \Big] \, .
\end{eqnarray}

\paragraph{Transformations to $C<0$:}
In order to obtain the above set of functions for negative values of the asymmetry constant, one has to derive the appropriate transformations, for instance,
\begin{eqnarray}
    w(r,C) = -e^{-\frac{2\pi C}{r_0}}
    \left[w(-r,-C) + \frac{3J}{2C(C^2+r_0^2)} (1-e^{\frac{2\pi C}{r_0}})\right] \ .
\end{eqnarray}

\section{Radial perturbations}

In order to obtain the perturbation equations for the radial perturbations, we start from the slowly rotating ($sr$) background discussed above.
We then perturb the fields linearly in first order in the perturbation parameter $\epsilon_q$.
The perturbed metric reads
\begin{eqnarray}
g_{\mu\nu} &=& g^{(sr)}_{\mu\nu} + \epsilon_q \delta h_{\mu\nu}(t,r,\theta,{\varphi}) \nonumber \\
 &=& g^{(sr)}_{\mu\nu} + \epsilon_q 
\left( \delta h^{(A)}_{\mu\nu}(t,r,\theta,{\varphi}) 
    +  \delta h^{(P)}_{\mu\nu}(t,r,\theta,{\varphi}) \right)
\, , 
\label{per_m}
\end{eqnarray}
with parity even, i.e., polar ($P$), and parity odd, i.e., axial ($A$), contributions, and the perturbed phantom field is given by
\begin{eqnarray}
\Phi &=& {\Phi}^{(sr)} + \epsilon_q \delta{\phi}^{(P)} (t,r,\theta,{\varphi})
\label{per_p}
\, .
\end{eqnarray}

Following closely the detailed derivations in \cite{Blazquez-Salcedo:2022eik}, where the general formalism was developed and then applied to Kerr-Newman black holes, we now decompose these perturbations in terms of spherical harmonics $Y[\mathrm{l},\mathrm{m}](\theta,{\varphi})$, where $\mathrm{l}$ and $\mathrm{m}$ denote the multipole numbers.
Rotation leads to an entanglement of even (polar) and odd (axial) perturbations, allowing for decoupling only in the static limit.
At the same time we obtain, in general, an infinite tower of equations labeled by the multipole number $\mathrm{l}$, coupling the perturbation functions in the expansions.
Only the multipole number $\mathrm{m}$ remains unaffected by rotation, i.e., different values of $\mathrm{m}$ are not coupled because of the axial symmetry of the rotating background solution.

Here, we will not construct the general set of modes of the rotating Ellis-Bronnikov wormholes.
Instead we are only interested in the so-called radial-led perturbations, that reduce in the static limit to the well-known unstable mode.
Thus we focus on $\mathrm{l}=0$, $\mathrm{m}=0$ polar-led perturbations. 
Since we restrict to second order in rotation, the infinite sum with respect to $\mathrm{l}$ is truncated already at $\mathrm{l}=1$.
Indeed, these $\mathrm{l}=0$, $\mathrm{m}=0$ polar-led  perturbations couple only to axial $\mathrm{l}=1$ perturbations \cite{Blazquez-Salcedo:2022eik}, simplifying the problem considerably.
Employing a harmonic time dependence with frequency $\omega$, the metric perturbations can then be parametrized as follows, 
\begin{eqnarray}
\label{met_pert}
\delta h_{\mu\nu}= e^{i\omega t}
    \begin{pmatrix}
NY_0 & 0    & 0 & {S}_0 Y_\theta \\
0   & LY_0 & 0 & {S}_1 Y_\theta \\
0 & 0 & R^2 TY_0 & 0 \\
{S}_0 Y_\theta & {S}_1 Y_\theta & 0 & R^2 T\sin^2{\theta}Y_0
\end{pmatrix} \, \, \, \, \, \, \, \, \, ,
\end{eqnarray}
where we have introduced the five radial functions $N, L, T, {S}_0, {S}_1$, and the abbreviations $Y_0=Y[0,0]=1/\sqrt{4\pi}$ and $Y_\theta=\sin{\theta} \, \partial_\theta Y[1,0] =-\sin^2{\theta}\sqrt{3/4\pi}$. 
A suitable parameterization of the phantom field perturbation is given by
\begin{eqnarray}
\label{phan_pert}
 \delta{\phi}^{(P)} = {e^{i\omega t}}{\phi_1(r)} Y_0 \, ,
\end{eqnarray}
with radial function $\phi_1(r)$.
Restricting to second order in the rotation parameter $\epsilon_r$, these perturbations lead to a consistent set of equations.

Insertion of the perturbations into the general set of field equations provides us with the set of ordinary differential equations for the six radial perturbation functions $N, L, T, {S}_0, {S}_1$ and $\phi_1(r)$ to be solved numerically.
To fix the gauge, we set ${S}_0$ and $\phi_1(r)=0$.
The remaining four unknown functions are then determined by solving a set of differential equations for $dN/dr, dL/dr, dT/dr$ and an algebraic relation for ${S}_1$.
In the following we present these equations for $C>0$,
\begin{eqnarray}
\frac{dL}{dr}
&=&
    \frac{12  { e}^{-2f}}
      { r {r_0} R^4 }
      \left(4r^{5}-10 {C} r^{4}+ 10 {C}^{2} r^{3} - 5  {C}^{3} r^{2}
      +2  ({Q_0^2} -{r_0^2} )^{2}  (10r 
      - {C} )  \right)^{-1}
      \nonumber
      \\ 
 & \times &
   \big\{   
    \frac{2}{3}    { e}^{3f} r R^2 T \frac{d \phi_{20}}{d r} 
    \big(
    Q_0^2 R^2
   \kappa_1 
     (\tan^{-1} (\frac{r}{{r_0}}) 
    -\frac{\pi}{2} \frac{d\phi_{20}}{d r}  ) 
     \nonumber
      \\ 
&    +& {r_0} (Q_0^2 
   \kappa_1 
    \phi_{20} (r)
    +r^{6}-3 {C} r^{5}
    +(4 {C}^{2}+\frac{r_0^{2}}{2}) r^{4}-(3 {C}^{2}+ r_0^{2}) Cr^{3}
     \nonumber
      \\ 
    &+& (21 Q_0^{4}-21 r_0^{4}-\frac{39}{4} {C}^{2} r_0^{2}) r^{2}
    -5 {C}r (Q_0^{4}-r_0^{4}-\frac{9}{20} {C}^{2} r_0^{2})  
     \nonumber
      \\ 
    &+&\frac{1}{2} ({Q_0^2} -{r_0^2} )^{2}  ({C}^{2}+r_0^{2}))  
    \big)
    -\frac{{ e}^{f} }{3}  \kappa_2 {r_0} R^4 
    (2 \omega^{2} 
    R^2 r T ({h_0} (r )-1/2 )  +L {b_0} (r )) 
     \nonumber
      \\ 
    &-& 4 { e}^{5f} r  {r_0} T J^{2} 
    (r^{4}-\frac{7 r^{3} {C}}{4}+\frac{9 {C}^{2} r^{2}}{8}-\frac{5 {C}^{3} r}{16}+\frac{({Q_0^2} -{r_0^2} )^{2}  }{2}) 
     \nonumber
      \\ 
    &-& \frac{1}{3} R^2 { e}^{2f}   
    \big( Q_0^2 r 
    R^2 \kappa_2  \frac{d\phi_{20}}{d r} 
     \frac{r_0f}{C}  L
     \nonumber
      \\ 
    &+& {r_0} (-4 {b_0} (r ) [r^{6}-\frac{11 {C} r^{5}}{4}+(\frac{r_0^{2}}{2}+\frac{29 {C}^{2}}{8}) r^{4}+(-\frac{3}{4} {C} r_0^{2}-\frac{45}{16} {C}^{3}) r^{3}
     \nonumber
      \\ 
    &+& (-\frac{41}{2} r_0^{4}
    +\frac{41}{2} Q_0^{4}-\frac{79}{8} {C}^{2} r_0^{2}) r^{2}-5 {C} (Q_0^{4}-r_0^{4}-\frac{39}{80} {C}^{2} r_0^{2}) r +2 ({Q_0^2} -{r_0^2} )^{3} ] T
     \nonumber
      \\ 
   & + & L r (Q_0^2 \kappa_2 \phi_{20} (r )-2 r^{6}
    +\frac{11 {C} r^{5}}{2}
    -(r_0^{2}+\frac{13 {C}^{2}}{2}) r^{4}+(\frac{17}{4} {C}^{3}+2 {C} r_0^{2}) r^{3}
         \nonumber
     \\ 
    &+& (-26 Q_0^{4}+26 r_0^{4}+\frac{23}{2} {C}^{2} r_0^{2}) r^{2}
    +\frac{1}{2} (11 {C} r (Q_0^{4}-r_0^{4}-\frac{9}{22} {C}^{2} r_0^{2}) )
      \nonumber
      \\ 
    &-& 2 Q_0^{6}+5 r_0^{2} Q_0^{4}-3 r_0^{6}-{C}^{2} r_0^{4}))  \big) 
    + \kappa_2 {r_0}  
    (J^{2} r  { e}^{4f} L +\frac{4}{3} \omega^{2} R^6  {b_0} (r ) T  ) 
    \big\} \, ,
    \label{dL_withc}
\end{eqnarray}
where $\kappa_1 = 
r^{4}-\frac{5 r^{3} {C}}{2}+\frac{9 {C}^{2} r^{2}}{4}-\frac{7 {C}^{3} r}{8}+2 ({Q_0^2} -{r_0^2} )^{2} $, and $\kappa_2 =   r^{4}-2 r^{3} {C} +\frac{3 {C}^{2} r^{2}}{2}-\frac{{C}^{3} r}{2}
    +({Q_0^2} -{r_0^2} )^{2}$,
\begin{eqnarray}
  \frac{dN}{dr}
  &=&
    -\frac{e^{-2f}}{(r-C/2)^3 rr_0R^2}
    \{ 
-\frac{e^{3f}}{4} rT
(r-C/2)
\big[
\frac{d\phi_{20}}{dr}
4Q_0^2 R^2 
r_0f
\nonumber
\\
&-& 8 r_0
\big(-\frac{1}{2} Q_0^2 \phi_{20}(r) C  
+
(r-C/2)(h_0(r)+1/2) (Cr+r_0^2)
\big)
\big]
\nonumber
\\
&+& 
(r-C/2) 
\big[
(r-C/2)
r_0 e^{f}
(r\omega^2R^4 T
+ b_0(r)
((-Cr+r^2-r_0^2) L
-2N R^2)
)
\nonumber 
\\
&+& 
\frac{re^{2f}}{4}
\big(
[
4 R^2 Q_0^2
(r-C/2) 
\frac{d\phi_{20}}{dr}
\frac{r_0f}{C}
 (L+2N)
\nonumber
\\
&+&
4 r_0 (
2b_0(r) R^2 T 
+ (r-C/2)
([Q_0^2\phi_{20}(r)
+(Cr+2r_0^2) (h_0(r)+1/2)
]L
\nonumber 
\\
&+ &
2 Q_0^2 \phi_{20}(r) N) 
)
]
\big)
+ 2 (r-C/2) r_0R^4 \omega^2
b_0(r) T
\big]
    \} 
    \nonumber 
    \\ 
    &+& 
    \frac{3e^{-2f} J^2}
    {2C^2 R^4 (r-C/2)^3  (C^2+r_0^2)^2}
      \nonumber 
    \\ 
    &  \times &
\big\{  3 {e}^{5f} T \big(r^{8}+\frac{3 {C} r^{7}}{2}+(2 {r_0}^{2}+\frac{{C}^{2}}{4}) r^{6}
    +(\frac{13}{3} {C} {r_0}^{2}+\frac{35}{24} {C}^{3}) r^{5}
    \nonumber
    \\
    &+&
    (\frac{1}{6} {C}^{2} r_0^{2}+\frac{7}{6} {C}^{4}-r_0^{4}
    +2 Q_0^{4}) r^{4}
    -\frac{{C}r^{3}} {3} 
    ({C}^{4}+\frac{33}{4} {C}^{2} r_0^{2}-24 Q_0^{4}+\frac{31}{2} r_0^{4})
     \nonumber 
    \\
    &+&(-\frac{8 {C}^{6}}{3}-\frac{67 {C}^{4} r_0^{2}}{6}+(-\frac{79 r_0^{4}}{4}+16 Q_0^{4}) {C}^{2}+4 r_0^{2} Q_0^{4}-4 r_0^{6}) r^{2}
     \nonumber 
    \\
    &-&\frac{{C}r}{6} ({C}^{6}+45 {C}^{4} r_0^{2}+(-96 Q_0^{4}+\frac{485 r_0^{4}}{4}) {C}^{2}-48 r_0^{2} Q_0^{4}+48 r_0^{6})-\frac{{C}^{8}}{6}
     \nonumber 
    \\
    &-& \frac{2 {C}^{6} r_0^{2}}{3}-4 r_0^{4} {C}^{4}+(8 r_0^{2} Q_0^{4}-9 r_0^{6}) {C}^{2}+2 r_0^{4} Q_0^{4}-2 r_0^{8}
    \big) 
     \nonumber 
    \\
   & -& 6 
     R^2 {e}^{3f} T
    \big(r^{6}-\frac{{C} r^{5}}{2}+(-\frac{3 {C}^{2}}{4}+r_0^{2}) r^{4}+(\frac{31}{24} {C}^{3}+\frac{1}{6} {C} r_0^{2}) r^{3}-(\frac{9}{8} {C}^{4}+\frac{7}{4} {C}^{2} r_0^{2}) r^{2}
     \nonumber 
    \\
    &+&(\frac{1}{2} {C}^{5}+\frac{9}{8} {C}^{3} r_0^{2}) r -\frac{{C}^{6}}{3}-2 {C}^{4} r_0^{2}+(4 Q_0^{4}-\frac{4}{3} r_0^{2} Q_0^{2}-\frac{11}{3} r_0^{4}) {C}^{2}
    +2 r_0^{2} Q_0^{4}-2 r_0^{6}
    \big) 
     \nonumber 
    \\
    &+& 3 R^4
    {e}^{f} T 
    (r^{4}-\frac{5}{2} r^{3} {C} +\frac{9}{4} {C}^{2} r^{2}-\frac{7}{8} {C}^{3} r -2 r_0^{4}-{C}^{2} r_0^{2}+2 Q_0^{4})
     \nonumber 
    \\
    &+&
    {e}^{4f}
    \big[(r^{8}+2 {C} r^{7}+(2 r_0^{2}
    +3 {C}^{2}/2) r^{6}
    +(4 {C} r_0^{2}
    +3 {C}^{3}/2) r^{5}
    +({C}^{2} r_0^{2}/2+Q_0^{4}) r^{4}
     \nonumber 
     \\
%     \end{eqnarray}
%    \newpage 
%\begin{eqnarray}
    &-& {C} ({C}^{4}+2 {C}^{2} r_0^{2}-4 Q_0^{4}+2 r_0^{4}) r^{3}+(-5 {C}^{6}/2
    - 17 {C}^{4} r_0^{2}/2
     \nonumber 
    \\
   & +& (-23 r_0^{4}/2+8 Q_0^{4}) {C}^{2}+2 r_0^{2} Q_0^{4}-2 r_0^{6}) r^{2}+{C} ({C}^{6}-2 {C}^{4} r_0^{2}
     \nonumber 
    \\
    &+& (8 Q_0^{4}-19 r_0^{4}/2 ) {C}^{2}+4 r_0^{2} Q_0^{4}-4 r_0^{6}) r -{C}^{8}/2
    -5 {C}^{6} r_0^{2}/2
    +(4 Q_0^{4}-6 r_0^{4}) {C}^{4}
     \nonumber 
    \\
    &+& (-9 r_0^{6} /2+4 r_0^{2} Q_0^{4}) {C}^{2}+r_0^{4} Q_0^{4}-r_0^{8}) L 
      \nonumber 
    \\
   & -& 4 {C} N ({C}^{2}+r_0^{2})
    (r^{3}+ 3 {C} r^{2}/2+({C}^{2}+r_0^{2}) r + {C} r_0^{2}/2 ) \left(r -C/2 \right)^{2}
    \big] 
     \nonumber 
    \\
    &-& 2R^2 \big[ {e}^{2f}
    \big(
    (r^{6}+(-{C}^{2}/2+r_0^{2}) r^{4}+{C}^{3} r^{3}/2+(-3 {C}^{2} r_0^{2}/2
    -15 {C}^{4}/16) r^{2}
     \nonumber 
    \\
    &+& ({C}^{3} r_0^{2}
    +5 {C}^{5}/8) r -C^6/4-{C}^{4} r_0^{2}+(-r_0^{2} Q_0^{2}+2 Q_0^{4}-\frac{3}{2} r_0^{4}) {C}^{2}+r_0^{2} Q_0^{4}-r_0^{6}) L
     \nonumber 
     \\
    &-& 2 {C} N ({C}^{2}+r_0^{2}) \left(r -C/2 \right)^{3}\big) 
     \nonumber 
    \\
    &-&
     \frac{1}{2}L R^2 (r^{4}-2 r^{3} {C} +\frac{3}{2} {C}^{2} r^{2}-\frac{1}{2} {C}^{3} r +Q_0^{4}-r_0^{4}-\frac{1}{2} {C}^{2} r_0^{2})
    \big]
\big\} \, ,
    \label{dN_withc}
\end{eqnarray}
\begin{eqnarray}
    \frac{dT}{dr}
    &=& \frac{3e^{2f}}{rr_0R^4(C/2 - r)} 
    \big\{ 
-\frac{2}{3} r_0R^2 e^{-3f}
(
(\frac{Cr^2}{2}
-\frac{r}{2} (\frac{C^2}{4}+r^2)
)L
+ \frac{R^2}{2}b_0(r)T
)
\nonumber
\\
&+&
\frac{2}{3}
e^{-4f} r_0R^2b_0(r) 
(Cr-r^2-C^2/4)
L
\nonumber
\\
&+& 
\frac{rR^2e^{-2f}}{3}
\big(
-\frac{Crr_0}{2}
+
R^2
Q_0^2
\frac{d\phi_{20}}{dr}
\frac{r_0 f}{C}
+r_0(Q_0^2\phi_{20}(r) + r^2)
\big)
T
\nonumber
\\
&+& J^2 r_0 r T
   \big\} \, ,
   \label{dT_withc}
\end{eqnarray}
and finally
\begin{eqnarray}
    S_1(r)
    &=& \frac{i\sqrt{3}e^{2f} J}
    {\omega CR^2 (C^2+r_0^2)}
    \big\{
    e^{-f}[
    \big(-\frac{1}{2}C^2(C+r)
    -\frac{R^2}{4}
    (2r+3C) \big) L
    + C(C^2+r_0^2)N
    ]
    \nonumber
    \\
    &+&
    \frac{1}{2}e^{-3f} R^2 (r-\frac{C}{2})  L
    +\big(R^2(r-C)e^{-2f}
    -2C^3
    -C(r^2+3r_0^2)
    -rR^2
    \big)T
    \big\}  \, .
    \nonumber \\
    \label{s1_withc}
\end{eqnarray}
The background functions $\phi_{20}(r)$, $b_0(r)$, and $h_0(r)$ 
are taken from equations (\ref{h0_withc}), (\ref{b0_withc}), (\ref{phi20_withc}), respectively.
In the symmetric case $C=0$ the equations (\ref{dL_withc}) - (\ref{s1_withc}) simplify.
They can be found in \cite{Azad:2023iju}.

We now show how to cast the above system of equations into a master equation, i.e., a single second order ordinary differential equation, with the remaining set of perturbation functions determined by algebraic relations.
To begin with we note that equation $dT/dr$ (\ref{dT_withc}) and equation $dL/dr$ (\ref{dL_withc}) are not coupled to the perturbation function $N$. 
We then take the derivative of equation $dT/dr$ (\ref{dT_withc}).
Next we make use of equations $dT/dr$ (\ref{dT_withc}) and $dL/dr$ (\ref{dL_withc}) to express the function $L$ and its derivative $dL/dr$ in terms of the function $T$ and its derivative $dT/dr$. 
In this way we obtain a second order equation for the function $T$ of the form
\begin{eqnarray}
\label{eqd2T}
\frac{d^2T}{dr^2}+ A(r)\frac{dT}{dr} + B(r)T = 0 \, ,
\end{eqnarray}
with the coefficient functions $A(r)$ and $B(r)$.
 
These coefficient functions $A(r)$ and $B(r)$ are obtained as combinations of the coefficients present in the perturbation equations $dT/dr$ (\ref{dT_withc}) and $dL/dr$ (\ref{dL_withc}). 
For convenience, we express these functions in the following form,
\begin{eqnarray}
    A(r) = \frac{r_*'' }{r_*'}+2\frac{\alpha'}{\alpha} \, , \\
    B(r) = \frac{\alpha'' }{\alpha}-\frac{r_*''\alpha' }{r_*'\alpha}-2\left(\frac{\alpha'}{\alpha}\right)^2+(r_*' )^2 (V-\omega^2)  \, ,
\end{eqnarray}
where we have introduced a tortoise coordinate $r_*$ and an amplitude $\alpha$.
The tortoise coordinate $r_*$ is defined by
\begin{eqnarray}
\label{tor_coord}
   \partial_r r_*
   = e^{-f}
   \left[
   1 +
   \epsilon_r^2 (e^{-f}b_0-h_0)
   \right] \, ,   
 \end{eqnarray}  
and the amplitude $\alpha$ by
\begin{eqnarray}
   \frac{\partial_r\alpha}{\alpha}
   = \frac{(\partial_r\phi)^2R^2
   +\epsilon_r^2
   [
   2e^{-f}b_0
   -
   e^{-2f}R^4(\partial_r w)^2/6
   ]
   }{r-\sqrt{R^2(\partial_r\phi)^2-r_0^2}}
   \, .  \, \, \, \, \, \, \, \, \, \, \, \, 
\end{eqnarray}
Here we have kept the expansion parameter $\epsilon_r$ to indicate the second order contributions.

For completeness, we now outline how to obtain the remaining perturbation functions, once the equation for the function $T$ has been solved.
Once the function $T$ and the derivative $dT/dr$ are known, we may express the perturbation function $L$ in terms of $T$ and $dT/dr$ via equation (\ref{dT_withc}).
Subsequently, we may obtain the function $N$ via equation (\ref{dN_withc}), which yields a relation of the form $N=-L-2T+\mathcal{O}(J^2)$.
Finally, we may obtain the function $S_1$ via equation (\ref{s1_withc}). 

Taking $T =\alpha Z$, the perturbation equation can be expressed as a single standard stationary Schrödinger-like equation with a potential $V(r)$,
\begin{eqnarray}
\label{schroe}
    \frac{d^2 Z}{d{r^2_*}}
    + (\omega^2 - V(r))Z = 0 \, ,
\end{eqnarray}
where the eigenvalue $\omega$ corresponds to the mode frequency.
For $C>0$ the potential $V(r)$ is given by
\begin{eqnarray}
    V(r) &=&
    \frac{e^{2f} }{4 R^8 (C-2r)^5    \alpha} 
  (C^4-8rC^3+4C (r^2-r_0^2) (3C-4r) - 16r_0^2 (3r^2+2r_0^2) )
  \nonumber 
  \\    
   &+&  J^2 \big\{
-408
R^{2} { e}^{\frac{2 C}{r_0}  (\tan^{-1}(\frac{r}{r_0}) +\tan^{-1}(\frac{C}{2 r_0}) -\pi)}
(C^{4}-\frac{2}{17} C^{2} r_0^{2}+\frac{8}{17} r_0^{4}) 
 \nonumber 
  \\  &&  \times 
[8 \gamma R^{2}  Q_0^{2} \tan^{-1}\left(\frac{r}{r_0}\right)
+ \frac{1}{4} C^{6} (-\frac{1}{2} R^{2} \pi 
+ r r_0 ) +  C^{5}
(\frac{1}{4} \pi  r R^{2}+\frac{5}{2} r_0^{3} -\frac{1}{2} r^{2} r_0 )
 \nonumber 
  \\   &&
+3  C^{4} (-\frac{1}{2} \pi R^4 + rr_0 (r^2 -4 r_0^{2}) )
 \nonumber 
  \\ &&   
+C^{3} (r^{5} \pi -\pi  r r_0^{4}-2 r^{4} r_0 +50 r_0^{3} r^{2}+12 r_0^{5}) 
 \nonumber 
  \\ &&   
+C^{2} (-6 \pi  r^{4} r_0^{2}-14 \pi  r^{2} r_0^{4}-8 \pi  r_0^{6}-68 r^{3} r_0^{3}-8 r_0^{5} r ) 
 \nonumber 
  \\ &&   
+4 r_0^{2} C (r^{5} \pi -\pi  r^{3} r_0^{2}-2 \pi  r r_0^{4}+12 r^{4} r_0 +8 r_0^{3} r^{2}+8 r_0^{5})  
 \nonumber 
  \\ &&   
-16 r_0^{3} (\pi  r^{2} r_0^{3}+\pi  \,r_0^{5}+r^{5}+2 r r_0^{2} R^2)
] 
 \nonumber 
 \\
%     \end{eqnarray}
%    \newpage 
%\begin{eqnarray}
&+& 96 Q_0^{2} \big[
{ e}^{4f} \alpha_1 
\big(
 -\frac{7 C^{10}}{4}+\frac{41 C^{9} r}{4}
-(25 r^{2}+\frac{35 r_0^{2}}{4}) C^{8}
+(\frac{103}{4} r^{3}+34 r \,r_0^{2}) C^{7}
 \nonumber 
  \\ &&   
-(\frac{57}{8} r^{4}+64 r^{2} r_0^{2}+\frac{151}{8} r_0^{4}) C^{6}
+(-\frac{15}{4} r^{5}
+\frac{97}{2} r^{3} r_0^{2}+\frac{153}{4} r \,r_0^{4}) C^{5}
 \nonumber 
  \\ &&   
-(\frac{133}{4} r_0^{2} r^{4}+\frac{77}{4} r_0^{6}+63 r^{2} r_0^{4}+\frac{19}{2} r^{6}) C^{4}+r (\frac{29}{2} r_0^{6}+\frac{45}{2} r_0^{2} r^{4}+36 r^{2} r_0^{4}+13r^{6}) C^{3}
 \nonumber 
  \\ &&   
-(6 r_0^{8}+31 r^{2} r_0^{6}+11 r^{6} r_0^{2}+36 r^{4} r_0^{4}) C^{2}+(24 r^{5} r_0^{4}+26 r^{3} r_0^{6}+12 r \,r_0^{8}+10 r^{7} r_0^{2}) C 
 \nonumber 
  \\ &&   
-8 r_0^{6} R^{4} 
\big) 
 \nonumber 
  \\ &&   
-
[10 R^{2} 
{ e}^{2f} (C^{2}+\frac{2 r_0^{2}}{5})
\big( 
4 R^2  Q_0^{2} 
\gamma
(\tan^{-1}\left(\frac{C}{2 r_0}\right) - \tan^{-1}\left(\frac{r}{r_0}\right))
 \nonumber 
  \\ &&   
-2  r_0  (r-\frac{C}{2}) 
(-\frac{C^{5} r}{8}+(\frac{r^{2}}{4}-r_0^{2}) C^{4}+(-\frac{3}{2} r^{3}+\frac{7}{2} r r_0^{2}) C^{3}+(r^{4}-13 r^{2} r_0^{2}-4 r_0^{4}) C^{2}
 \nonumber 
  \\ &&   
+10 r^{3} r_0^{2} C 
-4 r_0^{2} 
(r^4+2r_0^2r^2+2r_0^4)
)
\big) 
] 
\big] 
     \big\}  \, ,
     \label{pot_withc}
\end{eqnarray}
where
\begin{eqnarray}
    \gamma = \frac{C^{4}}{8}-\frac{C^{3} r}{4}+(\frac{3 r^{2}}{2}+r_0^{2}) C^{2}
-(r^{3}-2 r \,r_0^{2}) C + 4 r_0^{4} \, .
\end{eqnarray}
For $C=0$ the potential is given in \cite{Azad:2023iju}.

\begin{figure}[h!]
		\centering
    		\includegraphics[width=0.32\textwidth,angle=-90]{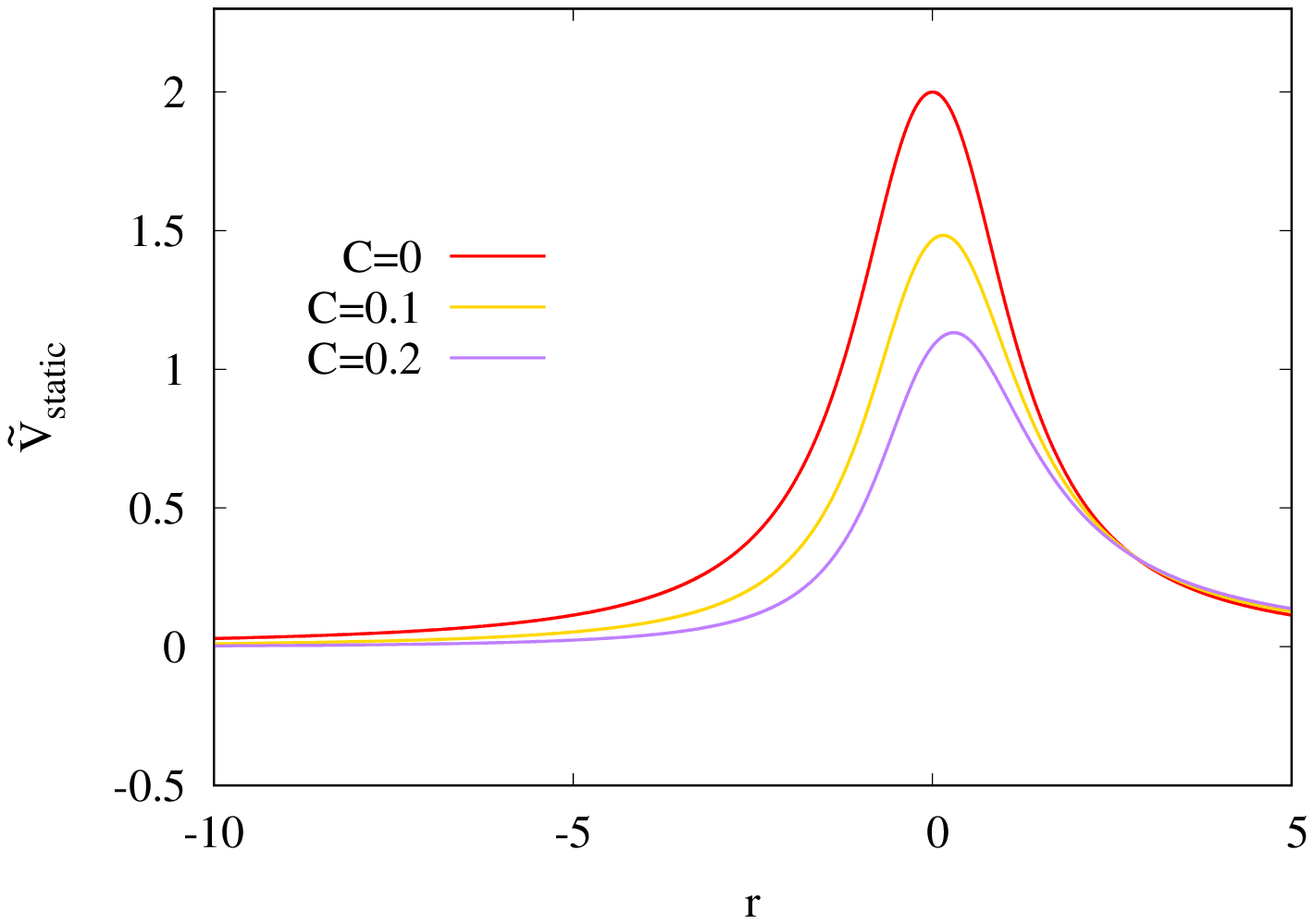}
\includegraphics[width=0.32\textwidth,angle=-90]{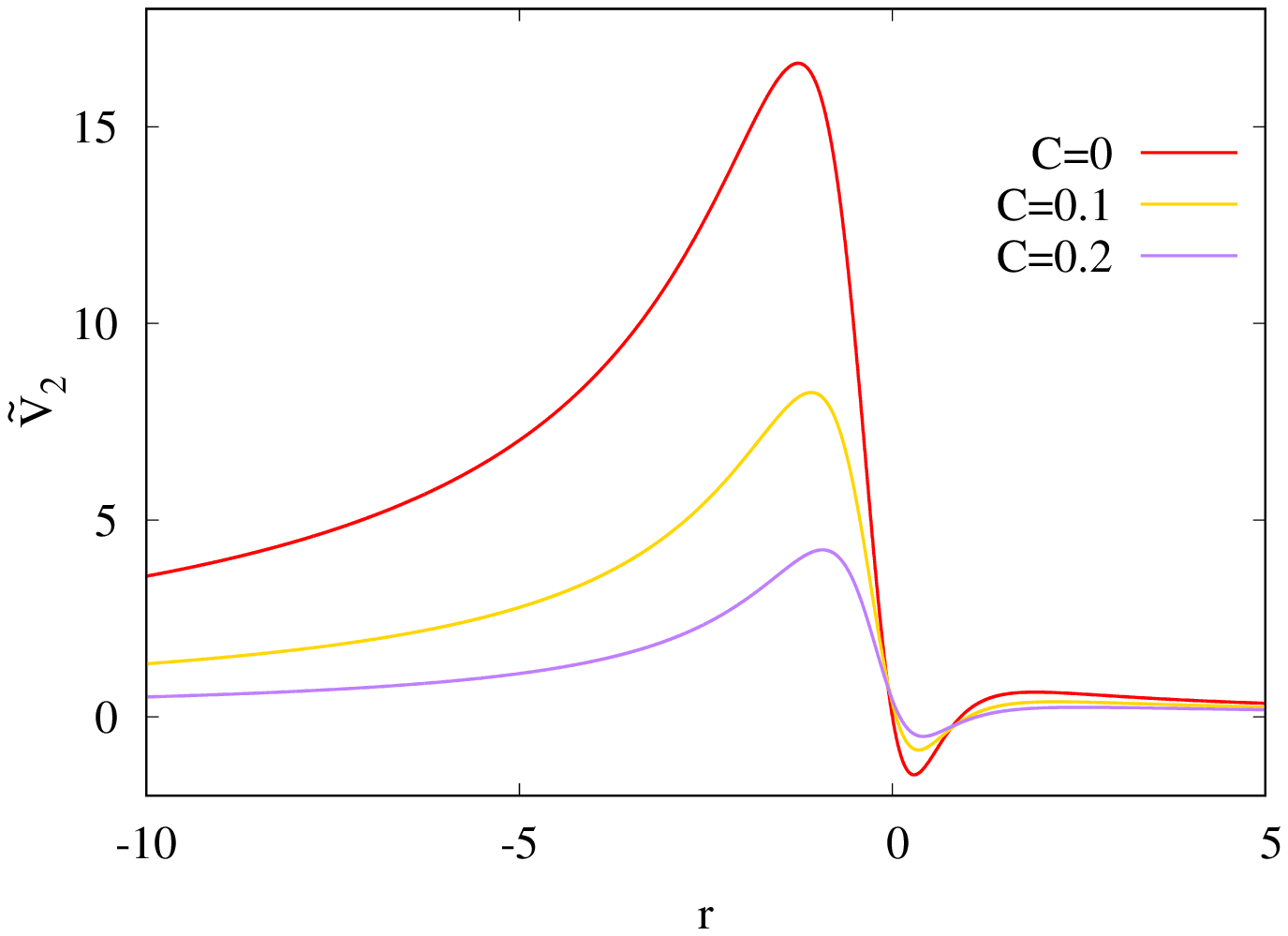}
		\caption{Static part $\tilde{V}_{static}=V_{static}(r-C/2)^2$ (left), and correction part $\tilde{V}_2=V_2(r-C/2)^2$ (right) of the potential vs radial coordinate $r$ for several values of the asymmetry parameter $C$ for $r_0=1$.}
		\label{Fig_pot}
	\end{figure}

The potential $V(r)$ consists of the static part $V_{static}(r)$ and the rotational correction term $V_2(r)$, entering $V(r)$ in second order
\begin{eqnarray}
    V(r) &=& V_{static}(r) + J^2 \, V_2(r) \, ,
    \label{pot_term}
\end{eqnarray}
with no linear term arising. 
We exhibit in Figure \ref{Fig_pot} the 
static term $V_{static}(r)$ and the
correction term $V_2(r)$ in equation (\ref{pot_term}), scaled with $\left(r-C/2\right)^2$, for several values of the asymmetry parameter $C$ (for throat parameter $r_0=1$). 

\section{Numerical method}

We now explain the numerical method we employ to obtain unstable modes.
For unstable modes the frequency $\omega$ entering via the harmonic time dependence of the perturbations becomes a purely imaginary number, $\omega = i\omega_I$ with $\omega_I<0$, leading to exponential growth in time. 
Thus we have to solve the Schr\"odinger-like equations (\ref{schroe}), looking for bound states. 
The boundary conditions for the perturbation function $T(r)$ obtained from an asymptotic analysis show, that $T(r)$ should decay exponentially at both asymptotic infinities, 
\begin{equation}
    \underset{r\to \pm \infty}{\lim} T(r) = 0 \, .
\end{equation}

We then compactify the radial coordinate according to 
\begin{equation}
    r = r_0 \tan(\pi x/2) \, .
\end{equation}
Thus the new radial coordinate takes values in the interval $-1 \le x \le +1$,  mapping the corresponding two radial infinities, $r=\pm \infty$, to the interval boundaries, $x=\pm 1$.
After performing this coordinate transformation we solve the Schr\"odinger-like equation for the function $T(x)$.
 
In fact, we solve this second order equation for $T(x)$ together with a first order equation, that is an auxiliary equation for the eigenvalue $E=\omega^2$
\begin{equation}
    E'=0 
    \, ,
\end{equation}
with the prime indicating differentiation with respect to $x$.
Since the equation for $T(x)$ is linear, and we are looking for a non-trivial solution, we choose an arbitrary point $x_c$ within the domain of integration, where we impose a finite value for the perturbation function $T(x)$.
Altogether, we then impose the following three boundary conditions
\begin{equation}
    T(\pm1)=0 , \ \ \ \ T(x_c)=1 \, .
\end{equation}
 
\begin{figure}[h!]
		\centering
    		\includegraphics[width=0.45\textwidth,angle=-90]{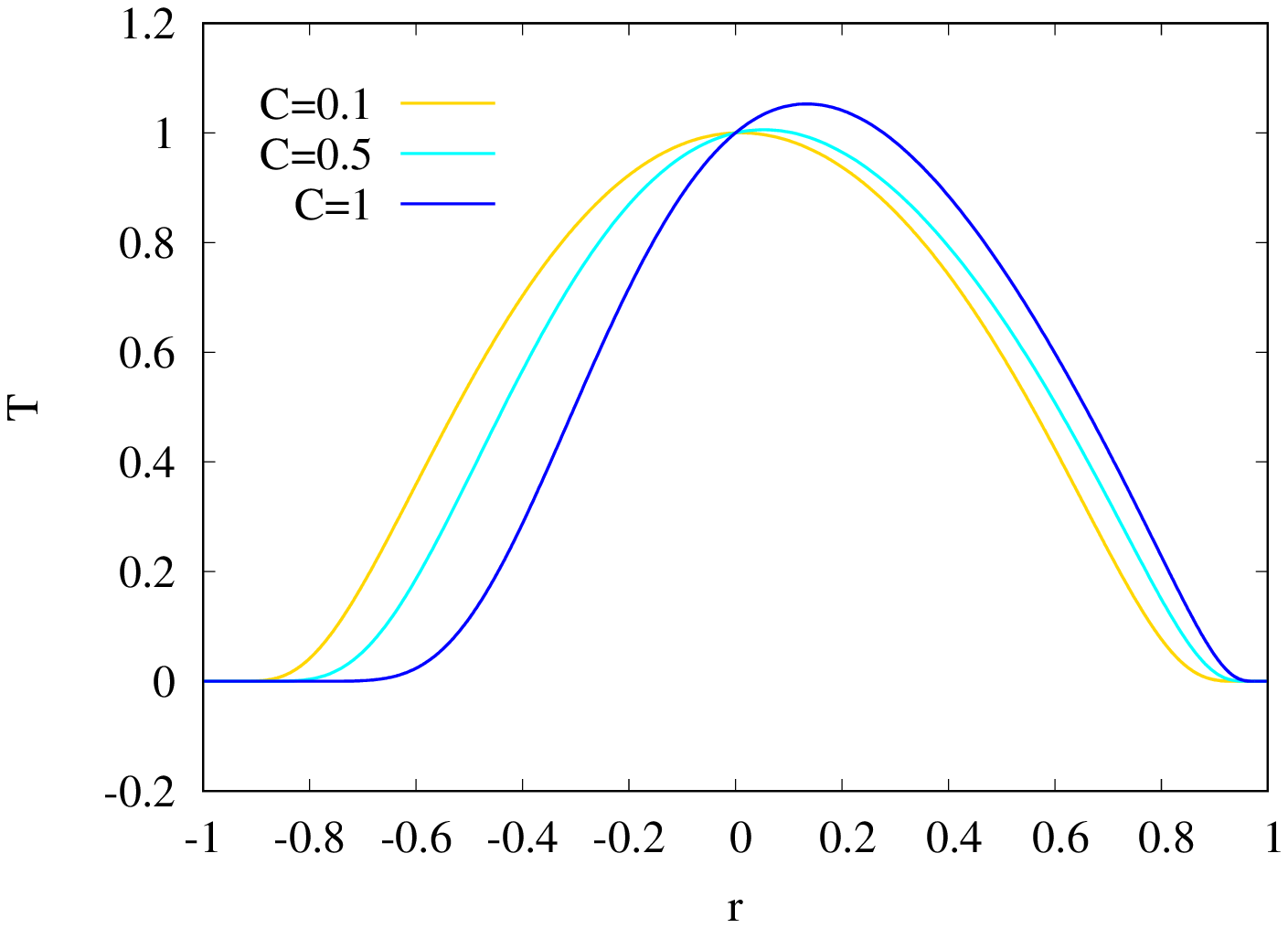}
		\caption{Perturbation function $T(r)$ {vs radial coordinate $r$} for several values of the asymmetry parameter $C$ for $r_0=1$.} 
		\label{Fig_pertT}
	\end{figure}

For the numerical calculation we employ the ordinary differential equation solver COLSYS \cite{Ascher:1979iha}.
COLSYS is a solver for boundary value problems that employs a spline collocation method, has automatic mesh adaptation, and calculates solutions with a user defined accuracy.
Since COLSYS uses a Newton-Raphson method it needs an initial guess.
As guess we give a Gaussian-like function for the perturbation function $T(x)$, satisfying the above boundary conditions, and a constant value for the eigenvalue $E(x)$ in the auxiliary equation.
In order to achieve convergence, a good guess for the eigenvalue is usually needed.
The typical accuracy of the solutions is on the order of $10^{-12}$.
We illustrate the solutions in Figure \ref{Fig_pertT}, where we show the perturbation function $T(r)$ for several values of the asymmetry parameter $C$.

In second order in rotation, we expect to obtain a quadratic dependence of the eigenvalue $\omega$ on the angular momentum $J$ of the rotating background configuration,
\begin{eqnarray}
    \omega_I = \omega_I^{(0)} + J^2 \Delta\omega_I^{(2)} \, ,
    \label{omega}
\end{eqnarray}
where the eigenvalue $\omega_I^{(0)}$ of the static background solution is corrected by the quadratic angular momentum term with coefficient $\Delta\omega_I^{(2)}$.
In order to determine this coefficient $\Delta\omega_I^{(2)}$ and thus the correction term of the unstable mode, we start from the static case $J=0$.
Employing fixed values for the asymmetry parameter $C$ and the wormhole parameter $r_0$, we then solve for the eigenvalue numerically for successively increasing values of the angular momentum $J$.

\section{Results}

\begin{figure}[h!]
		\centering
    		\includegraphics[width=0.6\textwidth,angle=-90]{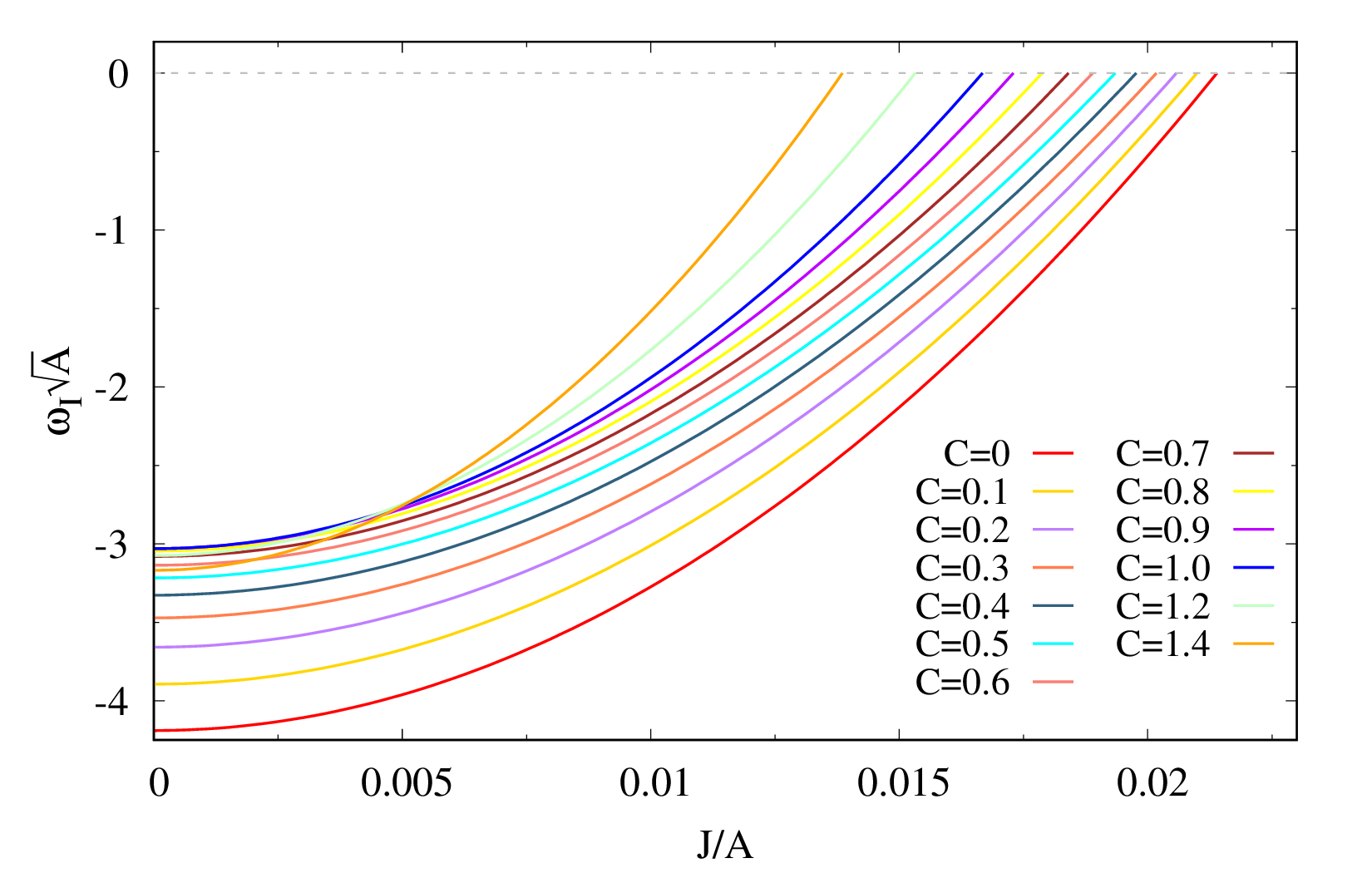}
  \caption{Scaled main unstable mode $\omega_I \sqrt{A}$ vs scaled angular momentum $J/A$ for several values of the asymmetry parameter $C$ and $r_0=1$. Scaling parameter is the throat area $A$.} 
		\label{Fig_omegaI_J}
	\end{figure}

\begin{table}[h!]
\begin{center}
\begin{tabular}{ || c | c c || }
 \hline
$C$ & $\omega^{(0)}_I$ & $\Delta\omega^{(2)}_I$ 
\\ 
  \hline
0 & -1.182 & 16.358
  \\ 
   0.1 & -1.016 & 10.663
  \\ 
  0.2 & -0.882 & 7.036
  \\ 
  0.3 & -0.774   &  4.691
  \\
  0.4 & -0.686 & 3.162
  \\ 
  0.5 &  -0.613   & 2.155
  \\
  0.6 & -0.552   & 1.488
  \\
  0.7 & -0.501 & 1.040
  \\ 
  0.8 & -0.458   & 0.736
  \\
  0.9 & -0.422   & 0.528
  \\ 
  1.0 & -0.390 & 0.384
  \\ 
  1.2 &  -0.338  & 0.210
  \\
  1.4 & -0.298    & 0.121
  \\ 
  \hline
\end{tabular}
\end{center}
\caption{Static term $\omega_I^{(0)}$ and correction term $\Delta \omega_I^{(2)}$ of the main unstable radial-led mode, eq.~(\ref{omega}), for a set of values of the asymmetry parameter $C$ in second-order in rotation.} 
\label{ImOm_coeff_main_moreC}
\end{table}

\begin{figure}[h!]
		\centering
    		\includegraphics[width=0.5\textwidth,angle=-90]{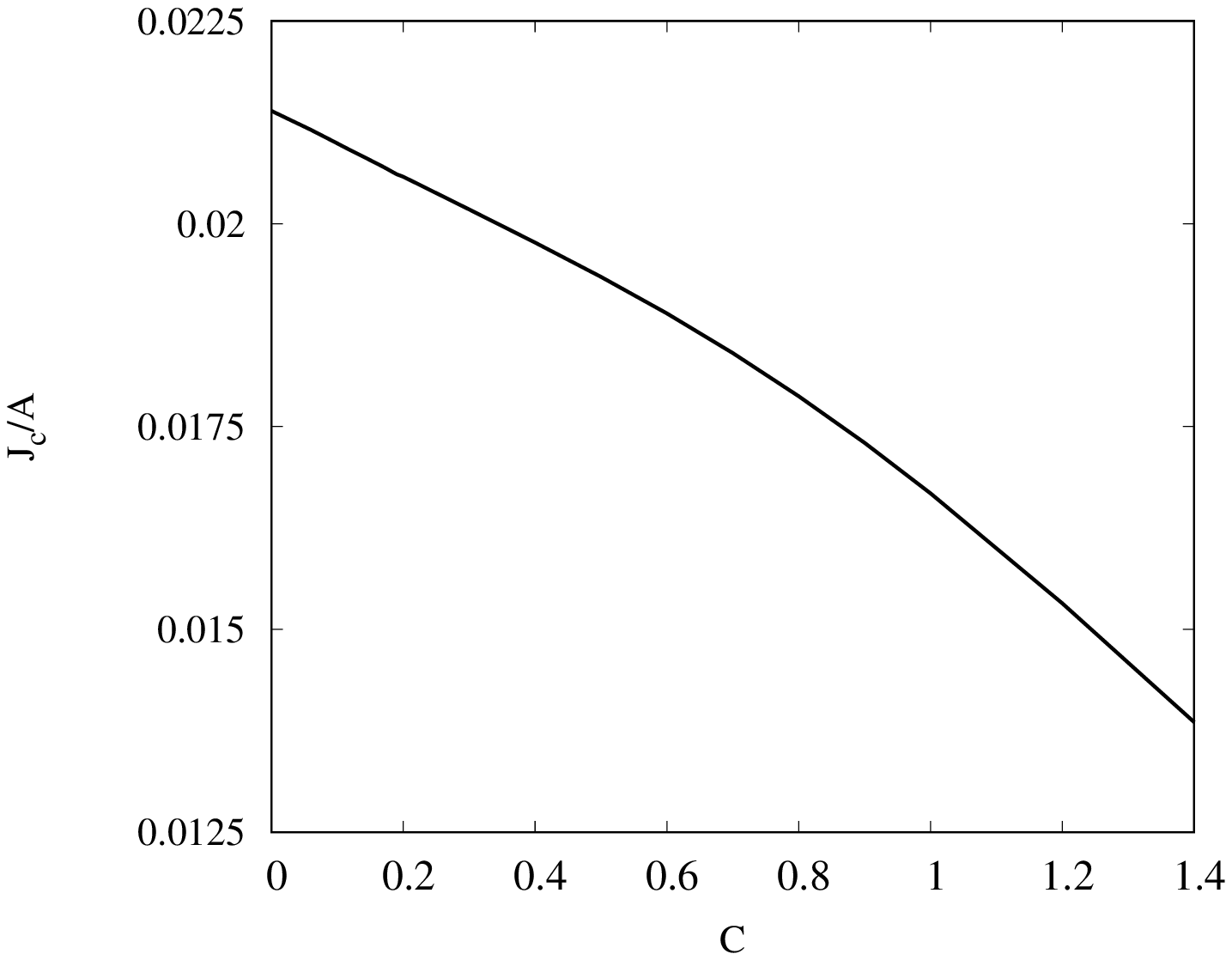}
		\caption{Critical angular momentum scaled with the throat area $J_c/A$ vs the asymmetry parameter $C$.} 
		\label{Fig_crit_J}
	\end{figure}

We now turn to the discussion of the numerical results, obtained in second order in rotation.
We first focus on the unstable mode, present already in the static case.
In the following we refer to this mode as the \textit{main} unstable mode.
In order to deal with dimensionless quantities, we employ the area of the  throat $A$ of the rotating background solutions, as a scaling parameter.
We illustrate the dependence of the main scaled unstable mode $\omega_I \sqrt{A}$ on the scaled angular momentum $J/A$ in Figure \ref{Fig_omegaI_J}, selecting a set of values of the asymmetry parameter $C$ and keeping $r_0=1$.

Figure \ref{Fig_omegaI_J} shows a monotonic increase of the scaled eigenvalue $\omega_I \sqrt{A}$ with increasing scaled angular momentum $J/A$, for all values of the asymmetry parameter $C$ considered.
In particular, for all values of $C$ the eigenvalue reaches zero at some critical value of the scaled angular momentum.
Denoting this critical value by $J_c/A$ we exhibit its dependence on the asymmetry parameter $C$ in Figure \ref{Fig_crit_J} for the range of values studied, where $J_c/A$ exhibits a monotonic decrease with increasing $C$.

To put these critical values into perspective we recall a crucial property of rapidly rotating Ellis-Bronnikov wormholes.
When the angular momentum $J$ of the wormholes is increased the associated set of wormholes approaches an extremal Kerr black hole for any given value of the asymmetry parameter $C$ \cite{Kleihaus:2014dla,Chew:2016epf}.
Extracting the scaled angular momentum $J/A$ (here $A$ denotes the horizon area of the extremal Kerr black hole) for this limiting solution yields $J/A\to 1/(8\pi) \approx 0.0397$.
Thus the critical angular momentum $J_c/A$ is about $50\%$ or less of the limiting value for the values of $C$ studied.

The monotonic increase of the eigenvalue $\omega_I$ with increasing angular momentum $J$, seen in Figure \ref{Fig_omegaI_J}, shows that all values of the second order correction $\Delta\omega^{(2)}_I$ are positive in the considered range of the asymmetry parameter $C$.
We exhibit the values of the static contribution $\omega^{(0)}_I$ and the associated correction $\Delta\omega^{(2)}_I$ entering the second order expansion for $\omega_I$ from equation (\ref{omega}) for a set of values of the asymmetry parameter $C$ in Table \ref{ImOm_coeff_main_moreC}.
Since the correction $\Delta\omega^{(2)}_I$ is positive, the mode becomes more stable with increasing angular momentum.
This indicates that rotation indeed can have a stabilizing effect on Ellis-Bronnikov wormholes. 
 
Currently, there are no higher order or exact numerical results available for 
the unstable mode of
rapidly rotating wormholes, to check the range of validity of the second order calculations.
However, the present method has been developed and applied before for Kerr and Kerr-Newman black holes \cite{Blazquez-Salcedo:2022eik}.
There it was shown that the second order results are excellent approximations of the exact results up to 50-60\% of the bound given by the extremal black holes.
By analogy we conclude, that the second order results for Ellis-Bronnikov wormholes should be excellent approximations for most of the range $J/A < J_c/A$ of the angular momentum.
 
We also remark that we have looked for the possible presence of a first order correction.
However, our numerical analysis showed, that such a first order correction is compatible with zero, in particular, it is always smaller than $10^{-7}$ in our calculations.
This is of course fully consistent within our formalism, since a linear correction would be proportional to the angular number $\mathrm{m}$, and $\mathrm{m}=0$ for the radial-led modes (see e.g. \cite{Blazquez-Salcedo:2022eik}).

  \begin{figure}[h!]
		\centering
    		\includegraphics[width=0.32\textwidth,angle=-90]
      {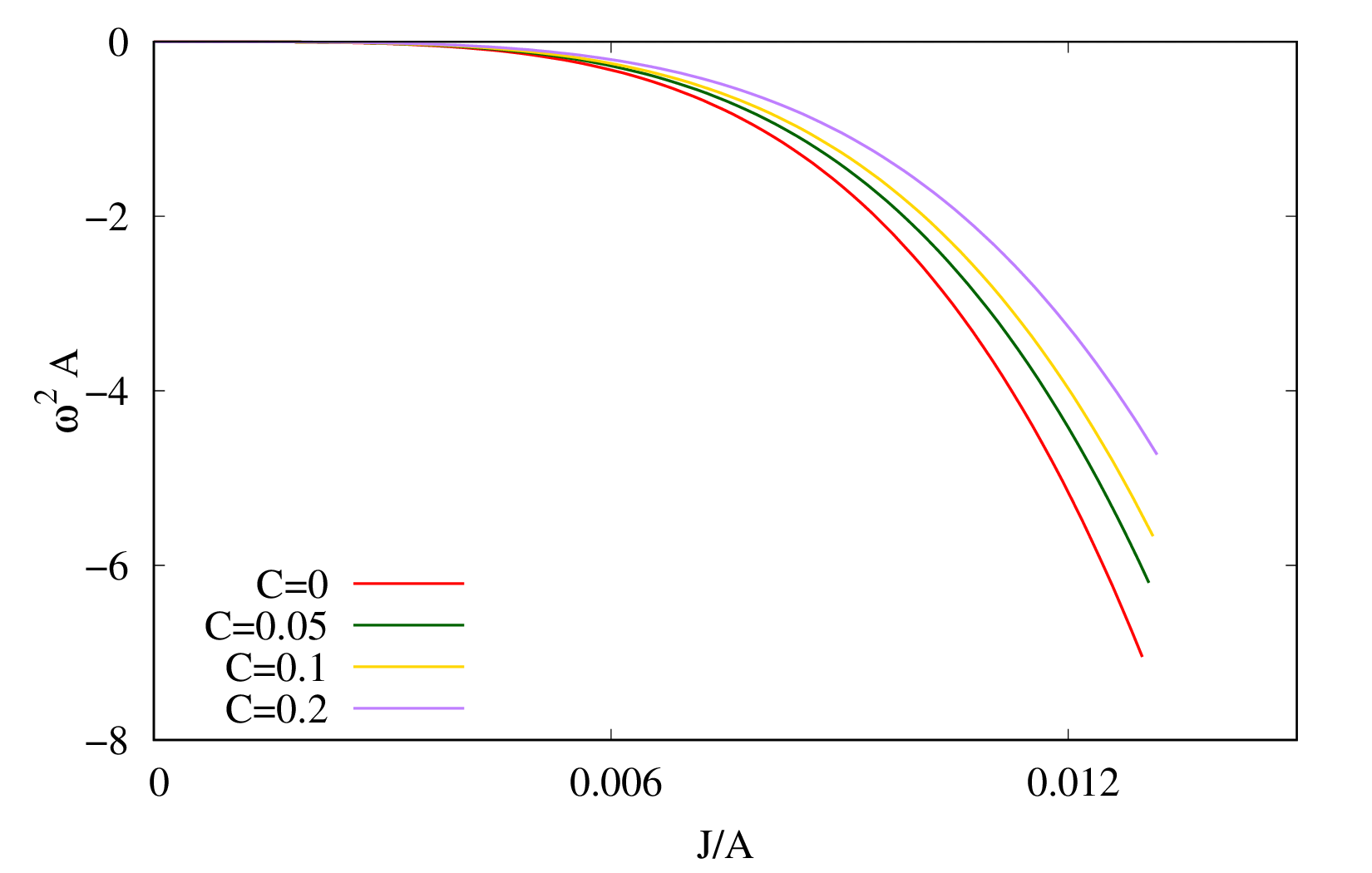}
		\caption{Second unstable mode vs scaled angular momentum $J/A$ for several values of the asymmetry parameter $C$.
        Scaling parameter is the throat area $A$.} 
		\label{Fig_J_w2}
	\end{figure}	

  \begin{figure}[h!]
		\centering
    		\includegraphics[width=0.32\textwidth,angle=-90]{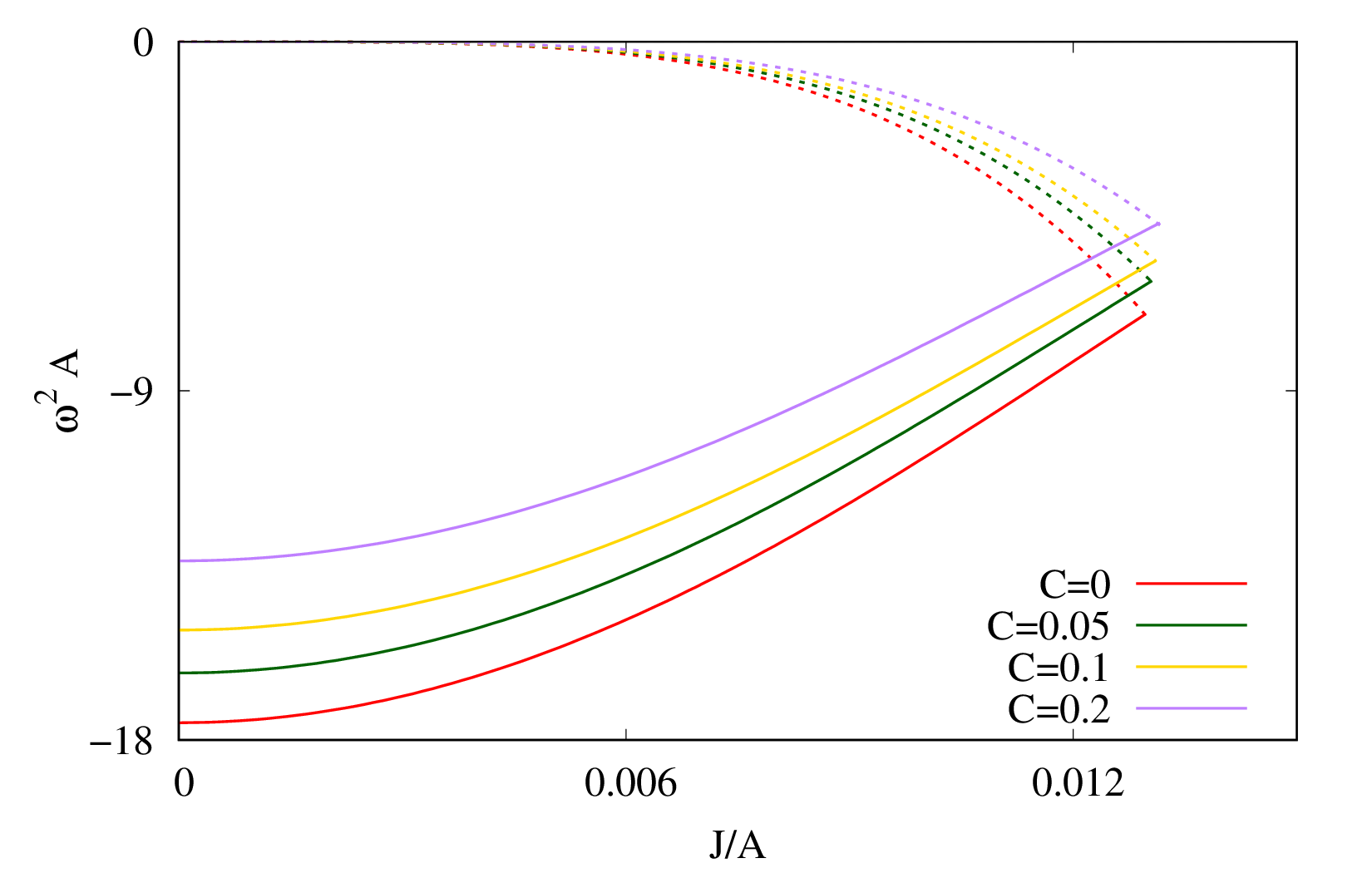}
		\caption{Scaled squared main unstable mode $\omega_I^2 A$ (solid) and second unstable mode (dotted) vs scaled angular momentum $J/A$ for $C=0$, $0.05$, $0.1$ and $0.2$.
        Scaling parameter is the throat area $A$.} 
		\label{Fig_J_w2_both}
	\end{figure}	

Besides the main unstable radial-led mode the slowly rotating Ellis-Bronnikov wormholes feature a second radial-led unstable mode.
This feature was first observed for rotating Ellis-Bronnikov wormholes in 5 dimensions, carrying equal magnitude angular momenta \cite{Dzhunushaliev:2013jja}.
As in higher dimensions \cite{Dzhunushaliev:2013jja}, here this second unstable mode arises also 
from a zero mode, present in the static limit. 
With increasing angular momentum of the wormholes this mode then becomes more unstable, i.e., $\omega_I$ decreases.
We exhibit this second unstable mode in Figure \ref{Fig_J_w2} for a set of values of the asymmetry parameter $C$ and $r_0=1$.

Clearly, for the second unstable mode the correction term $\omega_I^{(2)}$ is negative. 
But since the correction term $\Delta\omega_I^{(2)}$ is rather small, this second unstable mode has a rather small magnitude for small angular momenta.
Table \ref{ImOm_coeff_sec} gives $\omega_I^{(2)}$ for several values of $C$.
Since, independent of $C$, this second unstable mode starts from a zero mode in the static limit, only $\Delta\omega_I^{(2)}$ is given in Table \ref{ImOm_coeff_sec}. 
We have also checked that the second unstable mode features no linear dependence.

\begin{table}[h!]
\begin{center}
\begin{tabular}{ || c | c  || }
 \hline
$C$ &  $\Delta\omega^{(2)}_I$ 
\\ 
  \hline
0  & -28.185
  \\ 
   0.05  & -21.431
  \\ 
   0.1 & -16.697
  \\ 
  0.2 & -10.220
  \\ 
  \hline
\end{tabular}
\end{center}
\caption{Correction term $\Delta \omega_I^{(2)}$ of the second unstable radial-led mode, eq.~(\ref{omega}), for a set of values of the asymmetry parameter $C$ in second-order in rotation.
}
\label{ImOm_coeff_sec}
\end{table}

Since the main unstable mode increases with increasing angular momentum while the second unstable mode decreases, both modes are bound to cross.
This is illustrated in Figure \ref{Fig_J_w2_both}, where both modes are shown for a set of values of the asymmetry parameter $C$.
The respective crossing points $J_\times /A$ necessarily occur at values of $J/A$ smaller than the respective critical values $J_c/A$.
In particular, for the values of $C$ shown in the figure the crossing takes place around $J/A\approx0.013$, which slightly exceeds $30\%$ of the limiting value of $J/A$.

For each value of the asymmetry parameter $C$, we have truncated both branches of unstable modes at the respective crossing point $J_\times /A$ in Figure \ref{Fig_J_w2_both}.
The reason for this truncation is, that we expect that the two branches merge at some special value $J_s/A$ of the angular momentum $J/A$ and disappear.
This is what is observed in the non-perturbative calculations in 5 dimensions \cite{Dzhunushaliev:2013jja}, and we conclude by analogy that this will also hold for the wormholes in 4 dimensions.
Consequently these rotating wormholes then no longer feature a radial instability for $J/A > J_s/A$.
 
We attribute the crossing of the branches that we observe in 4 dimensions to a deficiency of the current perturbative approach, and expect to see a smooth merging of the modes also in 4 dimensions in a non-perturbative study.
Clearly, the applied quadratic approximation excludes a smooth merging of the branches, since this would require higher order terms in the angular momentum.
To verify the conjectured behavior a newly developed method might be employed, that is based on the spectral method \cite{Blazquez-Salcedo:2023hwg,Khoo:2024yeh}.

\section{Conclusions}

In this work we have investigated the influence of rotation on the radial instability of Ellis-Bronnikov wormholes.
The presence of this instability in static wormholes had been noticed already long ago \cite{Shinkai:2002gv,Gonzalez:2008wd,Gonzalez:2008xk} and the possibility of a stabilization by rotation had been suggested \cite{Matos:2005uh}.
Indeed, the disappearance of the radial instability with sufficiently rapid rotation of Ellis-Bronnikov wormholes could later be demonstrated in 5 dimensions for the special case of equal magnitude angular momenta \cite{Dzhunushaliev:2013jja}.

Here we have presented a detailed derivation and demonstration of the dependence of the radial instability of Ellis-Bronnikov wormholes in 4 dimensions in the case of slow rotation.
We have resorted to this perturbative study since the angular dependence does not factorize in 4 dimensions, making the study of the linear mode stability of rotating wormholes much more challenging than in the special 5-dimensional case.
Still, our study demonstrates great similarity of the dependence of the radial instability on the angular momentum of these wormholes in 4 and 5 dimensions. 

Working in second order in rotation, the radial polar $\mathrm{l}=0,\, \mathrm{m}=0$ perturbations of the static case acquire a contribution from the axial $\mathrm{l}=1,\, \mathrm{m}=0$ perturbations, representing thus radial-led perturbations.
Solving the resulting Schr\"odinger-like master equation for the modes then yields a quadratic dependence of the eigenvalues on the angular momentum of the wormholes.

Our numerical analysis reveals a positive correction term to the static eigenvalue for the unstable radial mode, present in the limiting static case, for all values of the asymmetry parameter of the wormholes considered.
Consequently, there arises always a critical angular momentum, when the mode reaches zero.
At the same time, a second unstable arises from a zero mode in the static limit, that decreases quadratically with increasing angular momentum.
Necessarily both modes cross at a value of the angular momentum below the critical value.

In  contrast, in 5 dimensions the two unstable modes smoothly merge at a critical value of the angular momentum and disappear beyond this critical value.
This is indeed the behavior, that we also conjecture to be present in 4 dimensions.
However, this behavior cannot be realized within the employed second order approximation, since in order to generate a smooth merging of the modes higher order terms in the angular momentum would be necessary.
Therefore we only conjecture here, that the true critical angular momentum, where the radial instability of the rotating Ellis-Bronnikov wormholes will disappear, is slightly smaller than the angular momentum, where the two perturbatively obtained branches of unstable radial modes cross.

A verification of this conjecture will need more sophisticated methods than perturbation theory.
Indeed, the radial-led modes will have to be calculated based on the exact (numerically known) rotating background solutions \cite{Kleihaus:2014dla,Chew:2016epf}.
Likewise, the determination of the modes will have to proceed by solving numerically the resulting systems of coupled partial differential equations.
Application of the recently developed spectral scheme \cite{Blazquez-Salcedo:2023hwg,Khoo:2024yeh} will hopefully provide the desired evidence.
However, as first calculations have already revealed, a different parametrization of the wormholes will probably be necessary.

\section*{Acknowledgement}
We gratefully acknowledge support by DAAD, DFG project Ku612/18-1, 
FCT project PTDC/FIS-AST/3041/2020, and MICINN project PID2021-125617NB-I00 ``QuasiMode".
JLBS gratefully acknowledges support from Santander-UCM project PR44/21‐29910.

\end{document}